\documentclass[leqno]{siamart190516}

\usepackage{url}
\usepackage{fdsymbol} 
\usepackage{graphicx, color}
\usepackage{enumitem}
\usepackage{hyperref}
\usepackage{xcolor}
\hypersetup{
    colorlinks,
    linkcolor={red!70!black},
    citecolor={blue!50!black},
    urlcolor={blue!80!black}
}
\usepackage[left=1in,right=1in,top=1in,bottom=1in,foot=0.5in]{geometry}

\usepackage{algpseudocodex}

\usepackage{tipa}
\usepackage[capitalize]{cleveref}

\newtheorem{claim}{Claim}

\theoremstyle{definition}

\newtheorem{remark}{Remark}

\newcommand{\op}{\ensuremath{{\rm op}}}

\newcommand{\M}{\mathcal{M}}
\newcommand{\wM}{\widehat{\mathcal{M}}}
\newcommand{\PP}{\mathcal{P}}
\newcommand{\C}{\mathcal{C}}
\newcommand{\RR}{\mathcal{R}}

\newcommand{\mm}{\ensuremath{{\rm mconv}}}

\newcommand{\co}[1]{\overline{#1}}

\newcommand\restr[2]{{
  \left.\kern-\nulldelimiterspace 
  #1 
  \vphantom{\big|} 
  \right|_{#2} 
  }}

\newlist{case}{enumerate}{3}
\setlist[case]{label=Case \arabic*:}
\setlist[case,1]{labelindent=\parindent,ref=\arabic*}
\setlist[case,2]{label=Case \arabic{casei}.\arabic*:,ref=\arabic{casei}.\arabic*}
\setlist[case,3]{label=Case \arabic{casei}.\arabic{caseii}.\arabic*:,ref=\arabic{casei}.\arabic{caseii}.\arabic*}

\DeclareMathOperator{\diam}{diam}
\DeclareMathOperator{\side}{side}
\DeclareMathOperator{\opposite}{opp}

\newcommand*{\emptylist}{[]}
\newcommand{\variable}[1]{\ensuremath{\mathit{#1}}}
\newcommand{\Let}{\State {\bfseries let} }

\renewcommand{\Return}{\State {\bfseries return} }
\algrenewcommand\algorithmicrequire{\textbf{Input:}}
\algrenewcommand\algorithmicensure{\textbf{Output:}}

\newcommand\mdoubleplus{\ensuremath{\mathbin{+\mkern-7mu+}}}
\renewcommand*{\append}{\mdoubleplus}
\newcommand*{\cons}{\cdot}

\newenvironment{runningexample}
{{\medskip\bfseries{The running example.}}}
{\medskip}

\title{Modules in Robinson Spaces}

\author{Mikhael Carmona\footnotemark[1] \footnotemark[2]
\and Victor Chepoi\thanks{LIS, Aix-Marseille Université, CNRS, and
Université de Toulon, Marseille, France (\email{\{mikhael.carmona, victor.chepoi, guyslain.naves,
pascal.prea\}@lis-lab.fr})}
\and Guyslain Naves\footnotemark[1]
\and Pascal Préa\footnotemark[1] \thanks{École Centrale Marseille, Marseille, France}
}

\headers{Modules in Robinson Spaces}{M. Carmona, V. Chepoi, G. Naves and P. Préa}

\date{}

\begin{document}

\maketitle

\begin{abstract}
  A \emph{Robinson space} is a dissimilarity space $(X,d)$ (i.e., a
  set $X$ of size $n$ and a dissimilarity $d$ on $X$) for which there
  exists a total order $<$ on $X$ such that $x<y<z$ implies that
  $d(x,z)\ge \max\{ d(x,y), d(y,z)\}$. Recognizing if a dissimilarity
  space is Robinson has numerous applications in seriation and
  classification. An \emph{mmodule} of $(X,d)$ (generalizing the
  notion of a module in graph theory) is a subset $M$ of $X$ which is
  not distinguishable from the outside of $M$, i.e., the distance from
  any point of $X\setminus M$ to all points of $M$ is the same.
  If $p$ is any point of $X$, then $\{ p\}$ and the maximal by inclusion
  mmodules of $(X,d)$ not containing $p$ define a partition of $X$,
  called the \emph{copoint partition}. In this paper, we investigate the
  structure of mmodules in Robinson
  spaces and  use it and the copoint partition to design a simple and practical
  divide-and-conquer algorithm for recognition of Robinson spaces in optimal
  $O(n^2)$ time.
\end{abstract}



\section{Introduction}

A major issue in classification and data analysis is to visualize
simple geometrical and relational structures between objects based on
their pairwise distances. 
Many applied algorithmic problems
ranging from archeological dating through DNA sequencing and numerical
ecology to sparse matrix reordering and overlapping clustering involve
ordering a set of objects so that closely coupled elements are placed
near each other. 
The classical {\it seriation problem},
introduced by Robinson \cite{Robinson} as a tool to seriate
archeological deposits, asks to find a simultaneous ordering (or
permutation) of the rows and the columns of the dissimilarity matrix
so that small values should be concentrated around the
main diagonal as closely as possible, whereas large values should fall
as far from it as possible. This goal is best achieved by considering
the so-called Robinson property: a dissimilarity matrix  
has the {\em Robinson property} if its values increase
monotonically in the rows and the columns when moving away from the
main diagonal in both directions. 
A {\it Robinson matrix} is
a 
dissimilarity matrix which can be transformed by permuting its rows
and columns to a 
matrix having the Robinson property. The permutation which leads to a
matrix with the Robinson property is called a {\it compatible order}.
Computing a compatible order can be viewed as the two-dimensional
version of the sorting problem.
In this paper, we present a simple and practical divide-and-conquer
algorithm for computing a compatible order and thus recognizing
Robinson matrices in optimal $O(n^2)$ time.



\subsection{Related work} 
%
Due to the importance 
in seriation and classification, the algorithmic problem of
recognizing Robinson dissimilarities/matrices 
attracted the interest of many authors and several polynomial time
recognition algorithms 
have been proposed. The existing 
algorithms can be classified into \emph{combinatorial} and
\emph{spectral}. All combinatorial algorithms are based on the
correspondence between Robinson dissimilarities and interval
hypergraphs. 
The main difficulty arising in recognition algorithms is the existence
of several compatible
orders. 
Historically, the first recognition algorithm was given in 1984 by
Mirkin and Rodin \cite{MiRo} and consists in testing if the hypergraph
of balls 
is an interval hypergraph; it runs in $O(n^4)$
time. 
Chepoi and Fichet \cite{ChepoiFichet} gave a simple divide-and-conquer
algorithm running in $O(n^{3})$
time. 
The algorithm
divides  the set of points into subsets and 
refines the
obtained subsets into classes to which the recursion can be applied.
Seston \cite{Seston} presented another $O(n^{3})$
time 
algorithm, by using threshold
graphs. 
In \cite{SestonThese}, he improved the complexity of his algorithm to
$O(n^2 \log n)$. 
 Finally, in 2014 Préa and Fortin
\cite{Prea} presented an algorithm running in optimal $O(n^2)$ time.
The efficiency of the algorithm of \cite{Prea} is due to the use of
the PQ-trees of Booth and Lueker \cite{BoothLueker} as a data
structure for encoding all compatible orders. Even if optimal, the
algorithm of \cite{Prea}
is far from being simple. Subsequently, two new recognition algorithms
were proposed by Laurent and Seminaroti: in \cite{LaSe1} they
presented an algorithm of complexity $O(\alpha\cdot n)$ based on
classical LexBFS traversal and divide-and-conquer (where $\alpha$ is
the depth of the recursion tree, which is at most the number of
distinct 
elements of the input matrix), and in \cite{LaSe2} they presented an
$O(n^2\log n)$ algorithm, which extends LexBFS to weighted matrices
and is used as a multisweep traversal. Laurent, Seminaroti and
Tanigawa \cite{LaSeTa} presented a 
characterization of Robinson matrices in terms of forbidden
substructures, extending the notion of asteroidal triples in graphs to
weighted graphs. More recently, Aracena and Thraves Caro \cite{ArThCa}
presented a parametrized algorithm for the NP-complete problem of
recognition of Robinson incomplete matrices ({\it i.e.} determining if an incomplete matrix can be completed into a Robinson one). Armstrong et al.
\cite{ArGuSiLo} presented an optimal $O(n^2)$ time algorithm for the
recognition of strict circular Robinson dissimilarities (Hubert et al.
\cite{HuArMe1} defined circular seriation first and it was studied
also in the papers \cite{ReKedA} and \cite{IsGiVe}). A simple and 
optimal algorithm for strict circular seriation was 
proposed in \cite{CaChNaPr_circ}.  

The spectral approach 
was introduced by Atkins et al. \cite{AtBoHe} and was subsequently
used in numerous papers (see, for example, \cite{FoDaVo} and the
references therein). The method is based on the computation of the
second smallest eigenvalue and its eigenvector of the Laplacian of a
similarity matrix $A$, 
called the {\em Fiedler
  value} and the {\it Fiedler vector} of $A$. Atkins et al.
\cite[Theorems 3.2 \& 3.3]{AtBoHe} proved that if 
$A$ is Robinson, then it has a monotone Fiedler vector and if $A$ is
Robinson with a Fiedler value and a Fiedler vector with no
repeated values, then the two permutations of the Fiedler vector in
which the coordinates are strictly increasing (respectively,
decreasing) are the only two compatible orders of $A$. 
For similarity matrices for which
the Fiedler vector has repeated values, Atkins et al. \cite{AtBoHe}
recursively apply the algorithm to each submatrix of $A$ defined by
coordinates of the Fiedler vector with the same value. In this case, they also use
PQ-trees to represent the compatible orders. This leads to an
algorithm of complexity $O(nT(n)+n^2\log n)$ to recognize if a
similarity matrix is Robinson, where $T(n)$ is the complexity of
computing the Fiedler vector of a matrix. The Fiedler vector is
computed by the Lanczos algorithm, which is an iterative numerical
algorithm that at each iteration performs a multiplication of the
input matrix $A$ by a vector. 

Real data contain errors, therefore the dissimilarity matrix $D$
can be measured only approximatively and 
fails to be Robinson. 
Thus 
we are lead to
the problem of approximating $D$ 
by a Robinson
dissimilarity $R$. As an error measure one can use the usual
$\ell_p$-distance $\Vert D-R\Vert_p$ between two $n\times n$
matrices. 
This $\ell_p$-fitting problem has been shown to be  NP-hard for $p=1$  \cite{BaBr} and for $p=\infty$ \cite{ChFiSe}. Various
heuristics for this optimization problem have been considered in
\cite{Br,Hu,HuArMe} and papers cited therein. The approximability of
this fitting problem for any $1\le p<\infty$ is open. Chepoi and
Seston \cite{ChSe} presented a factor 16 approximation for the
$\ell_{\infty}$-fitting problem. 
For a similarity matrix $A$, Ghandehari and Janssen
 \cite{GhJa1} 
introduced a parameter $\Gamma_1(A)$
and showed that one can construct 
a Robinson similarity $R$ (with the same order of lines and columns as $A$) such
that $\Vert A-R\Vert_1\le 26\Gamma_1(A)^{\frac{1}{3}}$. 
The result of Atkins et al. \cite{AtBoHe} in the case of the Fiedler
vectors with no repeated values was generalized by Fogel et
al. 
\cite{FoDaVo} to the case when the entries of
$A$ 
are subject to a uniform noise or some entries are not given. 
Basic examples of Robinson dissimilarities are the
ultrametrics. Similarly to the classical bijection
between ultrametrics and hierarchies, there is a one-to-one
correspondence between Robinson dissimilarities and pseudo-hierarchies due to Diday \cite{Di}
and Durand and Fichet \cite{DuFi}. Pseudo-hierarchies are classical structures in classification
with overlapping classes.



\subsection{Paper's organization}

The paper is organized as follows. The main notions related to the
Robinson property are given in Section~\ref{preliminaries}. In
Section~\ref{s:mmodules}, we introduce mmodules and copoints of a
dissimilarity space $(X,d)$ and present their basic properties. In
particular, we show that all copoints of a given point $p$ define a
partition of $X\setminus \{ p\}.$ In Section~\ref{s:flat-conical}, we
investigate the copoint partitions and the compatible orders for flat
and conical Robinson spaces. In Section~\ref{s:mmodulesRobinson}, we
investigate the properties of copoint partitions and their (extended)
quotients in general Robinson spaces. In Section~\ref{s:flat}, we
introduce the concept of proximity pre-order for an (unknown)
compatible order. 
We show that for extended quotients this pre-order is an order and we
show how to retrieve a compatible order from this proximity order. The concepts and the
results of Sections~\ref{s:mmodules}-\ref{s:flat} are used in the divide-and-conquer
algorithm, described and analyzed in
Section~\ref{s:divide-and-conquer-copoint}. 


\section{Preliminaries}\label{preliminaries}

In this section, we give some definitions which will be used
throughout this paper. When not defined just before their first use,
all notions and notations in this paper are defined here.

\subsection{Robinson dissimilarities}

Let $X=\{ p_1,\ldots, p_n\}$ be a set of $n$ elements, called
\emph{points}. A {\it dissimilarity} on $X$ is a symmetric function
$d$ from $X^2$ to the nonnegative real numbers such that
$d(x,y)=d(y,x)\ge 0$ and $d(x,y)=0$ if $x=y$. Then $d(x,y)$ is called
the {\it distance} between $x,y$ and $(X,d)$ is called a
\emph{dissimilarity space}. A partial order on $X$ is called
\emph{total} if any two elements of $X$ are comparable. Since we will
mainly deal with total orders, we abbreviately call them
\emph{orders}.

\begin{definition}[Robinson space]\label{def:Robinson}
  A dissimilarity $d$ and an order $<$ on $X$ are called {\em
    compatible} if $x<y<z$ implies that
  $d(x,z)\geq \mbox{max}\{ d(x,y),d(y,z)\}.$ If a dissimilarity space
  $(X,d)$ admits a compatible order, then $d$ is said to be {\em
    Robinson} and $(X,d)$ is called a {\em Robinson space}.
\end{definition}

Equivalently, $(X,d)$ is Robinson if its
distance matrix $D=(d(p_i,p_j))$ can be symmetrically permuted so that
its elements do not decrease when moving away from the main diagonal
along any row or column. Such a dissimilarity matrix $D$ is said to
have the {\it Robinson property} \cite{CrFi,Di,DuFi,Hu}. If
$Y\subset X$, we denote by $(Y,d|_{Y})$ (or simply by $(Y,d)$) the
dissimilarity space obtained by restricting $d$ to $Y$; we call
$(Y,d)$ a \emph{subspace} of $(X,d)$. If $(X,d)$ is a Robinson space,
then any subspace $(Y,d)$ of $(X,d)$ is also Robinson and the
restriction of any compatible order $<$ of $X$ to $Y$ is compatible.
If $d$ and $<$ are compatible, then $d$ is also compatible with the order $<^{\op}$
opposite to $<$.
Given two dissimilarity spaces $(X',d')$ and $(X,d)$, a map
$\varphi: X'\rightarrow X$ is an \emph{isometric embedding} of
$(X',d')$ in $(X,d)$ if for any $x,y\in X'$ we have
$d(\varphi(x),\varphi(y))=d'(x,y)$, i.e., if $(X',d')$ can be viewed
as a subspace of $(X,d)$.

The \emph{ball} of radius $r\ge 0$ centered at $x \in X$ is the set
$B_r(x):=\{y\in X: d(x,y)\leq r\}$. The \emph{diameter} of a subset
$Y$ of $X$ is $\diam(Y):=\max\{ d(x,y): x,y\in Y\}$ and a pair
$x,y\in Y$ such that $d(x,y)=\diam(Y)$ is called a \emph{diametral
  pair} of $Y$. A point $x$ of $Y$ is called \emph{non-diametral} if
$x$ does not belong to a diametral pair of $Y$. From the definition of
a Robinson dissimilarity follows that $d$ is Robinson if and only if
there exists an order $<$ on $X$ such that all balls $B_r(x)$ of
$(X,d)$ are intervals of $<$. Moreover, this property holds for all
compatible orders. Basic examples of Robinson dissimilarities are the
ultrametrics, thoroughly used in phylogeny. Recall, that $d$ is an
{\it ultrametric} if $d(x,y)\le \mbox{max}\{ d(x,z),d(y,z)\}$ for all
$x,y,z\in X$. Another example of a Robinson space is provided by the
standard {\it line-distance} between $n$ points $p_1<\ldots <p_n$ of
${\mathbb R}$. Any line-distance has exactly two compatible orders:
the order $p_1<\ldots<p_n$ defined by the coordinates of the points
and its opposite.

\begin{definition}[Flat spaces]\label{def:flat}
If a Robinson space $(X,d)$ has only two compatible
orders $<$ and $<^{op}$, then $(X,d)$ is said to be {\em flat}.
\end{definition}

Line-distances are flat but the converse is not true. We
conclude with the definition of conical spaces:

\begin{definition}[Conical spaces]\label{def:conical}
A dissimilarity
space $(X,d)$ is called \emph{conical} with \emph{apex} $p$ if all points of
$X\setminus \{ p\}$ have the same distance $\delta>0$ to $p$, i.e., $d(p,x)=\delta$
for any $x\in X\setminus \{p\}$. Since $p$ has the same distance $\delta$ to all points of
$X\setminus \{ p\}$, $(X,d)$ is a \emph{cone} over
$(X\setminus \{ p\},d)$ with apex $p$.
\end{definition}

\subsection{Algorithms and data structures}

Our algorithms will be written in pseudocode. They do not use any
fancy data structures besides lists and balanced binary search trees.

We use a bracketed notation for \emph{lists} with $[]$ being the empty
list. As a choice of presentation, we will use lists in a persistent
(or non-destructive) way~\cite{Ok}, meaning that a list cannot be
modified once defined. To this end, we introduce the two operators
$\cons$ and $\append$ defined by
\begin{align*}
x \cons [l_1,\ldots,l_n] &= [x,l_1,\ldots,l_n]\\
[l_1,\ldots,l_n] \append [l'_1,\ldots,l'_m] &= [l_1,\ldots,l_n,l'_1,\ldots,l'_m].
\end{align*}
One can implement the operator $\cons$ in $O(1)$ time and $\append$ in
$O(n)$ time ($n$ is the length of the left operand), using
single-linked lists. Extracting the first element of a list also takes
$O(1)$ time. We will also use the $\textrm{reverse}$ operation with
time-complexity $O(n)$, and $\textrm{concatenate}$ with
time-complexity $\sum_{i=1}^{k-1} |L_i|$, where:
\begin{align*}
\textrm{reverse}([l_1,l_2,\ldots,l_n]) &= [l_n, l_{n-1}, \ldots, l_1]\\
\textrm{concatenate}(L_1,\ldots,L_k) &= L_1 \append \ldots \append L_k.
\end{align*}

\emph{Balanced binary search trees} (see e.g. \cite{Ok}) are used
solely to sort in increasing order a list of $n$ elements with at most
$k$ distinct key values, in time $O(n \log k)$. This is achieved by
building a balanced binary search tree of the key values appearing in
the list, each associated to a list of elements sharing that key
value. The sorting algorithm consists in inserting each element in the
list associated to its key value, then concatenating all the
associated lists in increasing order of the key values. Each insertion
takes time $O(\log k)$, and the final concatenation takes time $O(n)$.
We denote the binary search tree operation by
$\textrm{insert}(T,\variable{key},\variable{value})$ (inserts a value
with a given comparable key), $\textrm{containsKey}(T,\variable{key})$
(checks whether there is a value with a given key),
$\textrm{get}(T,\variable{key})$ (retrieves the value associated to a
given key), and $\textrm{values}(T)$ (returns the list of keys in
increasing order of their values).

\subsection{Partitions and pre-orders}

A \emph{partition} of a set $X$ is a
family of sets $\PP=\{ B_1,\ldots,B_m\}$ such that
$B_i\cap B_j=\varnothing$ for any $i\ne j$ and
$\bigcup_{i=1}^k B_i=X$. The sets $B_1,\ldots, B_m$ are called the
\emph{classes} of $\PP$. A \emph{pre-order} is a partial order $\prec$
on $X$ for which incomparability is transitive. A partial order
$\prec$ on $X$ is a pre-order exactly when there exists an ordered
partition $\RR=(B_1,\ldots,B_m)$ of $X$ such that for $x\in B_i$ and
$y\in B_j$ we have $x\prec y$ if and only if $i<j$. Consequently, we
will also view a pre-order $\prec$ as an ordered partition
$\RR=(B_1,\ldots,B_m)$.
A partial order $\prec'$ \emph{extends} a partial order $\prec$ 
if $x\prec y$ implies $x\prec' y$ for all $x,y\in X$.

\begin{definition}[Stable partition]\label{def:stable-partition}
  A partition $\PP=\{ B_1,\ldots,B_m\}$ of a dissimilarity space $(X,d)$ is  a
  \emph{stable partition} if for any $i\ne j$ and for any three points
  $x,y\in B_i$ and $z\in B_j$, we have $d(z,x)=d(z,y)$.
\end{definition}

A non-stable partition $\PP$ can transformed into a stable partition by applying
the classical operation of \emph{partition refinement}, which proceeds as
follows. The algorithm maintains the current partition $\PP$ and for
each class $B$ of $\PP$ maintains the set $Z(B)$ of all points outside
$B$ which still have to be processed to refine $B$. While $\PP$
contains a class $B$ with nonempty $Z(B)$, the algorithm pick any
point $z$ of $Z(B)$ and partition $B$ into maximal classes that are
not distinguishable from $z$: i.e., for any such new class $B'$ and
any $x,x'\in B'$ we have $d(x,z)=d(x',z)$. Finally, the algorithm
removes $B$ from $\PP$ and insert each new class $B'$ in $\PP$ and
sets $Z(B'):=(B\setminus B')\cup (Z(B)\setminus \{ z\})$. Notice that
each class $B$ is partitioned into subclasses by comparing the
distances of points of $B$ to the point $z\notin B$ and such distances
never occur in later comparisons. Also, if the final stable partition
has classes $B'_1,\ldots, B'_t$, then the distances between points in
the same class $B'_i$ are never compared to other distances. This
algorithm is formalized in \Cref{algo:partitionRefine,algo:refine},
where one would call $\textrm{partitionRefine}(B, X \setminus B)$ for
each $B \in \PP$ to get a stable partition. We will not use
\Cref{algo:partitionRefine}, but will introduce and fully analyze a similar algorithm
\Cref{algo:recursiveRefine} that returns an ordered partition. It also uses \Cref{algo:refine}.
So, we now establish the complexity and correctness of \Cref{algo:refine}.


\begin{algorithm}[htbp]
  \caption{$\textrm{partitionRefine}(B,Z(B))$}
  \label{algo:partitionRefine}
  \begin{algorithmic}[1]
  \Require{a dissimilarity space $(X,d)$ (implicit),
    a class $B \subseteq X$ and a set $Z(B) \subseteq X\setminus B$}
  \Ensure{a partition $\{B_1,B_2,\ldots,B_k\}$ of $B$}
  \If{$Z(B) = \emptyset$}
      \Return $\{ B \}$
  \EndIf
  \Let $q \in Z(B)$,    \Comment{choose $q$ to be the first element of $Z(B)$}
  \Let $\{B_1,\ldots,B_m\} = \textrm{refine}(q,S)$ \Comment{ignore the order of the $B_i$s}
  \For {$i \in \{1,\ldots,m\}$}
      \Let $\PP_i = \textrm{partitionRefine}(B_i,
      \textrm{concatenate}(B_1, \ldots, B_{i-1},
                           B_{i+1}, \ldots, B_m, Z(B) \setminus \{q\}))$
  \EndFor
  \Return $\bigcup_{i=1}^m \PP_i$
  \end{algorithmic}
\end{algorithm}

\begin{algorithm}[htbp]
  \caption{$\textrm{refine}(q,S)$}
  \label{algo:refine}
  \begin{algorithmic}[1]
    \Require{a dissimilarity space $(X,d)$, 
      a point $q \in X$, a subset $S \subseteq X$.}
    \Ensure{an ordered partition of $S$, by increasing distance from $q$}
    \Let $T$ be an empty balanced binary tree, with keys in $\mathbb{N}$
    \For {$x \in S$}
      \If {$\lnot \textrm{containsKey}(T,d(q,x))$}
        \State $\textrm{insert}(T,d(q,x),[])$
      \EndIf
      \State $\textrm{insert}(T,d(q,x), x \cons \textrm{get}(T,d(q,x)))$
    \EndFor
    \Return $\textrm{values}(T)$
  \end{algorithmic}
\end{algorithm}

\begin{lemma}\label{lemma:refine-analysis}
  \Cref{algo:refine} called on $(q,S)$ outputs a partition $\mathcal{S} = (S_1,\ldots,S_m)$ of $S$ in  $O(|S| \log m)$ time, where
  \begin{enumerate}[label={\itshape (\roman*)}]
  \item for each $1 \leq i \leq m$, for all $x,y \in S_i$, $d(q,x) = d(q,y)$,
  \item for each $1 \leq i < j \leq m$, for all $x \in S_i, y \in S_j$, $d(q,x) < d(q,y)$.
  \end{enumerate}
\end{lemma}

\begin{proof}
  First, $\mathcal{S}$ is a partition, since each element is inserted in a
  list of $T$ exactly once. Each class of $\mathcal{S}$ is at constant distance from $q$
  since we use the distances to $q$ as keys. Finally, the classes of $\mathcal{S}$ are
  sorted by increasing distances from $q$, because $\textrm{values}(T)$
  returns its associated values in increasing order of keys.
  The complexity analysis follows from the fact that the binary search
  tree contains at most $m$ keys, hence each of its elementary
  operations are in $O(\log m)$. The evaluation of
  $\textrm{values}(T)$ can be done in $O(m)$ operations by a simple
  right-to-left DFS traversal of the binary search tree, inserting
  (not appending) each list in $T$ from farthest to closest into the
  returned list.
\end{proof}


\subsection{The running example} \label{section:runningExample}

Throughout the paper, we will use the dissimilarity space in
\Cref{TABLE_gros_example} and some of its subspaces to illustrate the
algorithms and the introduced notions. As will be seen in the final
\Cref{TABLE_gros_example_compatible}, this dissimilarity space is
Robinson, with the compatible order (among others):
$19 < 5 < 15 < 2 < 12 < 13 < 14 < 11 < 4 < 3 < 18 < 8 < 16 < 9 < 1 < 17 < 10 < 6 < 7.$

\begin{figure}[htbp]
{\scriptsize{
$$
\begin{array}{cp{0.1cm}ccccccccccccccccccc}
     & &  1 &  2 &  3 &  4 &  5 &  6 &  7 &  8 &  9 & 10 & 11 & 12 & 13 & 14 & 15 & 16 & 17 & 18 & 19 \vspace{0.1cm}\\
   1 & &  0 & 10 &  9 &  9 & 10 &  8 &  9 &  9 &  4 &  8 &  9 & 10 &  9 &  9 & 10 &  9 &  4 &  9 & 10 \\
   2 & &    &  0 &  8 &  8 &  2 & 10 & 11 &  8 & 10 & 10 &  5 &  2 &  5 &  5 &  1 &  8 & 10 &  8 &  2 \\
   3 & &    &    &  0 &  1 &  8 &  9 &  9 &  2 &  9 &  9 &  6 &  8 &  6 &  6 &  8 &  2 &  9 &  2 &  8 \\
   4 & &    &    &    &  0 &  8 &  9 &  9 &  2 &  9 &  9 &  6 &  8 &  6 &  6 &  8 &  3 &  9 &  2 &  8 \\
   5 & &    &    &    &    &  0 & 10 & 11 &  8 & 10 & 10 &  5 &  3 &  5 &  5 &  2 &  8 & 10 &  8 &  1 \\
   6 & &    &    &    &    &    &  0 &  9 &  9 &  9 &  7 &  9 & 10 &  9 &  9 & 10 &  9 &  8 &  9 & 10 \\
   7 & &    &    &    &    &    &    &  0 &  9 &  9 &  9 &  9 & 11 &  9 &  9 & 11 &  9 &  9 &  9 & 11 \\
   8 & &    &    &    &    &    &    &    &  0 &  9 &  9 &  6 &  8 &  6 &  6 &  8 &  2 &  9 &  2 &  8 \\
   9 & &    &    &    &    &    &    &    &    &  0 &  8 &  9 & 10 &  9 &  9 & 10 &  9 &  6 &  9 & 10 \\
  10 & &    &    &    &    &    &    &    &    &    &  0 &  9 & 10 &  9 &  9 & 10 &  9 &  7 &  9 & 10 \\
  11 & &    &    &    &    &    &    &    &    &    &    &  0 &  5 &  1 &  1 &  5 &  6 &  9 &  6 &  5 \\
  12 & &    &    &    &    &    &    &    &    &    &    &    &  0 &  5 &  5 &  2 &  8 & 10 &  8 &  4 \\
  13 & &    &    &    &    &    &    &    &    &    &    &    &    &  0 &  1 &  5 &  6 &  9 &  6 &  5 \\
  14 & &    &    &    &    &    &    &    &    &    &    &    &    &    &  0 &  5 &  6 &  9 &  6 &  5 \\
  15 & &    &    &    &    &    &    &    &    &    &    &    &    &    &    &  0 &  8 & 10 &  8 &  2 \\
  16 & &    &    &    &    &    &    &    &    &    &    &    &    &    &    &    &  0 &  9 &  2 &  8 \\
  17 & &    &    &    &    &    &    &    &    &    &    &    &    &    &    &    &    &  0 &  9 & 10 \\
  18 & &    &    &    &    &    &    &    &    &    &    &    &    &    &    &    &    &    &  0 &  8 \\
  19 & &    &    &    &    &    &    &    &    &    &    &    &    &    &    &    &    &    &    &  0 \\
\end{array}
$$
}}
\caption{\small{A distance matrix $D$ of a Robinson space $(X,d)$ with $X = \{1,\ldots, 19\}$.}
\label{TABLE_gros_example}}
\end{figure}

To illustrate the notions of flat and conical subspaces, notice that
the subspace $\{5, 14, 3, 9, 7\}$ is flat, with compatible orders
$5 < 14 < 3 < 9 < 7$ and its reverse. This follows from the fact that
$\{5,7\}$ is the unique diametral pair, whence 5 and 7 must be the
extremities of any compatible order. Then sorting the remaining points
by their distances from 5 imposes the rest of the order. One can
check in \Cref{TABLE_flat_and_conical}. Notice also that
the subspace $\{1,6,7,9,10\}$ is conical with apex 7 and $\delta = 9$.

The stable partition algorithm applied to the partition
$\{ \{2,5\}, \{1,3,4,6,7,8\} \}$ will return the partition
$\{\{2,5\}, \{3,4,8\}, \{1,6,7\}\}$. This is done by using $2$ as a
pivot on $\{1,3,4,5,7,8\}$, because $d(2,\{3,4,8\}) = 8$ while
$d(2,\{1,6\}) = 10$ and $d(2,\{7\}) = 11$. One can check that this
partition is stable, see \Cref{TABLE_flat_and_conical}.

\begin{figure}[htbp]
  \begin{center}\scriptsize
    \begin{equation*}
      \begin{array}{cccccc}
        &  5 & 14 &  3 &  9 &  7 \vspace{.1cm} \\
        5 &  0 &  5 &  8 & 10 & 11 \\
        14 &    &  0 &  6 &  9 &  9 \\
        3 &    &    &  0 &  9 &  9 \\
        9 &    &    &    &  0 &  9 \\
        7 &    &    &    &    &  0 \\
      \end{array} \qquad\qquad
      \begin{array}{cccccc}
        &  1 &  6 &  7 &  9 & 10 \vspace{.1cm} \\
        1 &  0 &  8 &  9 &  4 &  8 \\
        6 &    &  0 &  9 &  9 &  7 \\
        7 &    &    &  0 &  9 &  9 \\
        9 &    &    &    &  0 &  8 \\
        10 &    &    &    &    &  0 \\
      \end{array}\qquad\qquad
      \begin{array}{ccc|ccc|cc|c}
        &  2 &  5 &  3 &  4 &  8 &  1 &  6 &  7 \,\,\, \\
        2 &  0 &  2 &  8 &  8 &  8 & 10 & 10 & 11^{\,^{\,^{\,}}} \\
        5 &    &  0 &  8 &  8 &  8 & 10 & 10 & 11 \,\,\, \\ \hline
        3 &    &    &  0 &  1 &  2 &  9 &  9 &  9 \,\,\, \\
        4 &    &    &    &  0 &  2 &  9 &  9 &  9 \,\,\, \\
        8 &    &    &    &    &  0 &  9 &  9 &  9 \,\,\, \\ \hline
        1 &    &    &    &    &    &  0 &  8 &  9 \,\,\, \\ 
        6 &    &    &    &    &    &    &  0 &  9 \,\,\, \\ \hline
        7 &    &    &    &    &    &    &    &  0 \,\,\, \\
      \end{array}
    \end{equation*}
  \end{center}
  \caption{An illustration of several subspaces of $(X,d)$, from left
    to right: a flat subspace, a conical subspace with apex $7$, a
    subspace with an explicit stable partition.}
\label{TABLE_flat_and_conical}
\end{figure}

%

\section{Mmodules in dissimilarity spaces}\label{s:mmodules}

In this section, we introduce and investigate the notion of mmodule.
As one can see from their use in this paper, our motivation for
introducing them stems from the property of classes in stable
partitions: the points of the same class $C$ cannot be distinguished
from the outside, i.e., for any two points $x,y\in C$ and any point
$z\notin C$, the equality $d(z,x)=d(z,y)$ holds. After having obtained
the main properties of mmodules in general dissimilarities presented
in \Cref{mmodulesgeneral}, we discovered that our mmodules
coincide with ``clans'' in symmetric 2-structures, defined and
investigated by Ehrenfeucht and Rozenberg \cite{Ehren_Rozen_1,
  Ehren_Rozen_2} (see also Chein, Habib and Maurer \cite{ChHaMa}).
Since their theory is developed in a more general non-symmetric
setting, we prefer to give a self-contained presentation of elementary
properties of mmodules. Applying an argument from abstract convexity,
we deduce that for each point $p$ all maximal by inclusion mmodules
not containing $p$ together with $p$ define a partition of $X$. This
copoint partition is used in our divide-and-conquer algorithm for
recognizing Robinson spaces.

\subsection{Mmodules}\label{mmodulesgeneral} We continue with the definition
of mmodule of a dissimilarity space $(X,d)$.

\begin{definition}[Mmodules] \label{def:mmodules} A set $M\subseteq X$ is called
an \emph{mmodule} (a \emph{metric module} or a \emph{matrix module},
pronounced \textipa{[Em 'm6dju:l]}) if for any $z\in X\setminus M$ and
all $x,y\in M$ we have $d(z,x)=d(z,y)$.
\end{definition}

In graph theory, the subgraphs
indistinguishable from the outside are called modules (see
\cite{EhGaMcCSu,HaPa}), explaining our choice of the term ``mmodule''.
Denote by $\M=\M(X,d)$ the set of all mmodules of $(X,d)$. Trivially,
$\varnothing, X,$ and $\{ p\}, p\in X$ are mmodules; we call them
\emph{trivial mmodules}. An mmodule $M$ is called \emph{maximal} if
$M$ is a maximal by inclusion mmodule different from $X$.

\begin{runningexample}
  The sets
  $\{1,6,9,10,17\}, \{2,5,12,15,19\}, \{3,4,8,16,18\}, \{7\}, \{13,14,15\}$
  are the maximal mmodules of the running example. The set
  $\{12,5,19\}$ is a non-maximal mmodule.
\end{runningexample}

We continue with the basic properties of mmodules.

\begin{proposition} \label{mmodules} The set $\M=\M(X,d)$ has the following properties:
  \begin{enumerate}[label=(\roman*),nosep]
  \item\label{item:mm1} $M_1,M_2\in \M$ implies that $M_1\cap M_2\in \M$;
  \item\label{item:mm2} if $M\in \M$ and $M'\subset M$, then $M'\in \M$ if and
only if $M'$ is an mmodule of $(M,d)$;
\item\label{item:mm3} if $M_1,M_2\in \M$ and
  $M_1\cap M_2 \ne \varnothing$, then $M_1\cup M_2 \in \M$, furthermore, if
  $M_1\setminus M_2 \ne \varnothing$ and
  $M_2\setminus M_1 \ne \varnothing$, then
  $M_1\setminus M_2, M_2\setminus M_1,M_1\bigtriangleup M_2\in \M$;
  \item\label{item:mm4} the union $M_1\cup M_2$ of two intersecting maximal
mmodules $M_1,M_2\in \M$ is $X$;
  \item\label{item:mm5} if $M_1$ and $M_2$ are two disjoint maximal mmodules and
$M$ is a nontrivial mmodule contained in $M_1\cup M_2$, then either
$M\subset M_1$ or $M\subset M_2$;
  \item\label{item:mm6} if $M_1,M_2\in \M$ and $M_1\cap M_2=\varnothing$, then
$d(u,v)=d(u',v')$ for any (not necessarily distinct) points
$u,u'\in M_1$ and $v,v'\in M_2$;
  \item\label{item:mm7} if $\M'$
    is any partition of $X$
    into mmodules, then $\M'$ is a stable partition.
  \end{enumerate}
\end{proposition}

\begin{proof}
  To~\ref{item:mm1}: Pick any $x\notin M_1\cap M_2$ and $u,v\in M_1\cap M_2$. If
  $x\notin M_1\cup M_2$, then $d(x,u)=d(x,v)$ since $M_1,M_2\in \M$.
  If $x\in M_1\cup M_2$, say $x\in M_2\setminus M_1$, then
  $d(x,u)=d(x,v)$ since $M_1\in \M$.

  To~\ref{item:mm2}: First, let $M'$ be an mmodule of $(X,d)$. This implies that
  $d(x,u)=d(x,v)$ for any $x\in M\setminus M'$ and $u,v\in M'$, thus
  $M'$ is an mmodule of $(M,d)$. Conversely, let $M'$ be an mmodule
  of $(M,d)$ and we assert that $M'$ is an mmodule of $(X,d)$. Pick
  any $x\in X\setminus M'$ and $u,v\in M'$. If $x\in X\setminus M$, then
  $d(x,u)=d(x,v)$ since $u,v\in M'\subset M$ and $M$ is an mmodule. If
  $x\in M\setminus M'$, then $d(x,u)=d(x,v)$ since $M'$ is an
  mmodule of $(M,d)$ and we are done.

  To~\ref{item:mm3}: We first show that $M_1\cup M_2 \in \mathcal{M}$.
  If $M_1\cup M_2=X$, we are done. Otherwise, pick any
  $x\in X\setminus (M_1\cup M_2)$ and $u,v\in M_1\cup M_2$. If
  $u,v\in M_1$ or $u,v\in M_2$, then $d(x,u)=d(x,v)$ because
  $M_1,M_2\in \M$. Thus, let $u\in M_1\setminus M_2$ and
  $v\in M_2\setminus M_1$. Pick any $w\in M_1\cap M_2$. Then
  $d(x,u)=d(x,w)$ and $d(x,v)=d(x,w)$ since $M_1$ and $M_2$ are
  mmodules. Consequently, $d(x,u)=d(x,v)$ and thus
  $M_1\cup M_2\in \M$.

  Since $M_1,M_2\in \M$, for any
  $u,v\in M_1\setminus M_2, u',v'\in M_2\setminus M_1$,
  $y\in M_1\cap M_2$, and $x\in X\setminus (M_1\cup M_2)$, we have
  $d(x,u)=d(x,v)=d(x,y)=d(x,u')=d(x,v')$ and
  $d(u,u')=d(v,v')=d(u,y)=d(v,y)=d(u',y)=d(v',y)$. This shows that
  $M_1\setminus M_2, M_2\setminus M_1,$ and $M_1\Delta M_2$ are
  mmodules.

  To~\ref{item:mm4}: This is a direct consequence of (iii) and the
  definition of maximal mmodules.


  To~\ref{item:mm5}: Since $M$ is nontrivial, if $M$ is not equal to
  one of the $M_i$'s ($i\in \{1,2\}$), then we have
  $\varnothing \neq M\cap M_i \subsetneq M_i$ for, say, $i=2$. If
  $M\not\subset M_2$, then $M\cap M_1 \neq \varnothing$ and thus,
  by~\ref{item:mm3}, $M\cup M_1$ is an mmodule which, as
  $M_1\subsetneq M\cup M_1 \neq X$, contradicts the maximality of
  $M_1$.

  To~\ref{item:mm6}: Since $M_2$ is an mmodule and $u\notin M_2$,
  $d(u,v)=d(u,v')$. Since $M_1$ is an mmodule and $v'\notin M_1$,
  $d(v',u)=d(v',u')$. Consequently, $d(u,v)=d(u',v')$.

  To~\ref{item:mm7}: This follows from the definition of mmodules and the fact
  that $\M'$ partitions $X$.
\end{proof}

By \Cref{mmodules}\ref{item:mm1}, $\M$ is closed by intersection,
thus $(X,\M)$ is a \emph{convexity structure} \cite{VdV}. Thus for
each subset $A$ of $X$ we can define the \emph{convex hull} $\mm(A)$
of $A$ as the smallest mmodule containing $A$: $\mm(A)$ is the
intersection of all mmodules containing $A$. For points $u,v\in X$, we
call $\langle u,v\rangle:=\{ x\in X: d(x,u)\ne d(x,v)\}$ the
\emph{interval} between $u$ and $v$.

\begin{lemma}\label{mint}
  $\langle u,v\rangle\subseteq \mm(u,v)$.
\end{lemma}

\begin{proof}
  Pick $x$ in $\langle u,v\rangle$. If $x\notin \mm(u,v)$, then $d(x,u')=d(x,v')$
  for any $u',v'\in \mm(u,v)$. This is impossible since
  $u,v\in \mm(u,v)$ and $d(x,u)\ne d(x,v)$ by the definition of
  $\langle u,v\rangle$.
\end{proof}

The converse inclusion is not true. However, the following lemma shows
that $\M$ is an \emph{interval convexity} \cite{VdV} in the following
sense:

\begin{lemma}\label{intconv}
  $A \subseteq X$ is an mmodule if and only if
  $\langle u,v \rangle \subseteq A$ for any two points $u,v\in A$.
\end{lemma}

\begin{proof}
  By \Cref{mint}, $\langle u,v \rangle \subseteq \mm(u,v)$. Since
  $\mm$ is a convexity operator, $\mm(u,v) \subseteq \mm(A)$. Thus, if
  $A\in \M$, then $\langle u,v \rangle \subseteq \mm(A)=A$. Conversely,
  suppose $\langle u,v \rangle \subseteq A$ for any $u,v\in A$. If $A$
  is not an mmodule, there exist $x \in S\setminus A$ and $u,v \in A$
  such that $d(x,u) \ne d(x,v)$. But this implies that $x$ belongs to
  $\langle u,v \rangle \subseteq A$, contrary to the choice of $x$.
\end{proof}

\subsection{Copoint partition} \label{Section:copoint-partition} We
continue by defining copoints. This term arises from abstract
convexity \cite{Ja,VdV}. Then we prove that the copoints attached to
any point $p$ of $(X,d)$ are pairwise disjoint.

\begin{definition}[Copoint]\label{def:copoints}
  A \emph{copoint at a point $p$} (or a \emph{$p$-copoint}) is any
  maximal by inclusion mmodule $C$ not containing $p$; the point $p$
  is the \emph{attaching point} of $C$.
\end{definition}

\begin{runningexample}
  The copoints of point $1$ are $C_1 = \{9\}$, $C_2 = \{17\}$,
  $C_3 = \{6\}$, $C_4 = \{10\}$, $C_5 = \{3,4,8,16,18\}$,
  $C_6 = \{7\}$, $C_7 = \{11,13,14\}$ and $C_8 = \{2,5,12,15,19\}$.
  This can be readily checked using \Cref{fig:matrix-copoints}. 
\end{runningexample}

The copoints of $\M$ minimally generate $\M$, in the sense that each
mmodule $M$ is the intersection of the copoints containing $M$
\cite{VdV}. Denote by $\C_p$ the set of all copoints at $p$ plus the
trivial mmodule $\{ p\}$.

\begin{lemma}\label{copoint-partition}
  For any $p\in X$, $\C_p$ defines a partition of $X$.
\end{lemma}

\begin{proof} Pick any copoints $C,C'$ at $p$. If
  $C\cap C'\ne \varnothing$, by \Cref{mmodules}\ref{item:mm3}, the
  union $C\cup C'$ is an mmodule not containing $p$, contrary to the
  assumption that $C,C'$ are copoints at $p$. Since for any point
  $q\ne p$, $\{q\}$ is an mmodule, $q$ is contained in a copoint at
  $p$. Thus $\C_p$ defines a partition of $X$.
\end{proof}

\begin{definition}[Copoint partition]\label{def:copoint-partition}
  Consequently, we call $\C_p:=\{ C_0:=\{ p\}, C_1,\ldots,C_k\}$ the
  \emph{copoint partition} of $(X,d)$ with attaching point $p$.
\end{definition}

From \Cref{mmodules}\ref{item:mm7} it follows that $\C_p$ is a stable
partition of $X$ (see \Cref{def:stable-partition}).
$\C_p$ can be constructed by applying
the stable partition algorithm to the initial partition
$\{ \{ p\},X\setminus\{ p\}\}$.

\begin{definition}[Trivial and cotrivial copoint partitions]\label{def:trivial-cotrivial}
  The copoint partition $\C_p$ is called \emph{trivial} if $\C_p$
  consists only of the points of $X$, i.e.,
  $\C_p=\{ \{ x\}: x\in X\}$, and \emph{cotrivial} if
  $\C_p=\{ \{ p\}, X\setminus \{ p\}\}$, i.e., all points of
  $X\setminus \{ p\}$ have the same distance to $p$. If $\C_p$ is
  trivial, then $(X,d)$ is called \emph{$p$-trivial}.
\end{definition}

Notice that the copoint partition $\C_p$ is cotrivial if and only if
$(X,d)$ is conical with apex $p$ (see \Cref{def:conical}).
The following result follows directly from the
definitions:

\begin{lemma}\label{trivial-co-trivial}
  For a dissimilarity space $(X,d)$, the following holds:
  \begin{itemize}
  \item[(i)] $(X,d)$ is $p$-trivial for all $p\in X$
  if and only if all mmodules of $(X,d)$ are trivial;
  \item[(ii)] $(X,d)$ is conical for all $p\in X$
  if and only if $d(x,y)=\delta$ for all $x\ne y$ and some  $\delta>0$;
  \item[(iii)] if $(X,d)$ is conical with apex $p$, then each
    mmodule of $(X\setminus \{ p\}, d)$ is an
    mmodule of $(X,d)$.
  \end{itemize}
\end{lemma}

The heart of our divide-and-conquer algorithm is a decomposition of
the dissimilarity $(X,d)$ into the dissimilarities of its copoints, on
which we recurse. The merge step on the other hand will use the
quotient space:

\begin{definition}[Quotient space]\label{def:quotient}
Let $\C_p=\{ C_0=\{ p\}, C_1,\ldots,C_k\}$. The \emph{quotient space}
$(\C_p,\widehat{d})$ of $(X,d)$ has the classes of $\C_p$ as points
and for $C_i,C_j, i\ne j$ of $\C_p$ we set
$\widehat{d}(C_i,C_j):=d(u,v)$ for an arbitrary pair
$u\in C_i,v\in C_j$.
\end{definition}

\begin{runningexample}
  Considering also $C_0 = \{1\}$, we get a quotient space
  $(\C_1,\widehat{d})$ given in \Cref{fig:quotient_space_2}.

  \begin{figure}[htbp]
    {\footnotesize{
        \begin{equation*}
          \begin{array}{ccccccccccc}
                & &C_0 &C_1 &C_2 &C_3 &C_4 &C_5 &C_6 &C_7 &C_8 \vspace{0.1cm}\\
            C_0 & &  0 &  4 &  4 &  8 &  8 &  9 &  9 &  9 & 10 \\
            C_1 & &    &  0 &  6 &  9 &  8 &  9 &  9 &  9 & 10 \\
            C_2 & &    &    &  0 &  8 &  7 &  9 &  9 &  9 & 10 \\
            C_3 & &    &    &    &  0 &  7 &  9 &  9 &  9 & 10 \\
            C_4 & &    &    &    &    &  0 &  9 &  9 &  9 & 10 \\
            C_5 & &    &    &    &    &    &  0 &  9 &  6 &  8 \\
            C_6 & &    &    &    &    &    &    &  0 &  9 & 11 \\
            C_7 & &    &    &    &    &    &    &    &  0 &  5 \\
            C_8 & &    &    &    &    &    &    &    &    &  0 \\
          \end{array}
        \end{equation*}
      }}
    \caption{The quotient space $(\C_1,\widehat{d})$}
    \label{fig:quotient_space_2}
  \end{figure}
\end{runningexample}

\begin{lemma}\label{rcd*}
  The quotient space $(\C_p,\widehat{d})$ is $C_0$-trivial.
\end{lemma}

\begin{proof}
  Let $\C_p = \{C_0=\{p\}, C_1,\ldots, C_k\}$ and suppose that
  $(\C_p,\widehat{d})$ has a non-trivial mmodule $M$ not containing
  $C_0$. For any $C_j, C_{j'} \in M$ and $C_i \in \C_p\setminus M$, we
  have $\widehat{d}(C_i, C_j) = \widehat{d}(C_i, C_{j'})$. Setting
  $Y := \bigcup_{M} C_i$, for any $x, y \in Y$ and
  $z \in X \setminus Y$, we have $d(z,x) = d(z,y)$. Consequently, $Y$
  is an mmodule of $(X,d)$ not containing $p$, contradicting the
  maximality of the $C_i$'s.
\end{proof}

The definition of the quotient space implies that one can permute the
rows and columns of $D$ to get the following nice property. Partition
the rows and the columns of the distance matrix $D$ of $(X,d)$ into
sets corresponding to the copoints of $\C_p$, and permute the rows and
the columns of $D$, starting with the rows and columns corresponding
to the first copoint $C_0=\{ p\}$ of $\M$, then to the second copoint
$C_1$ of $\C_p$, etc. Then, in the resulting permuted matrix $D'$, for
each pair $C_i,C_j, i\ne j,$ of $\C_p$, the entries of $D'$
corresponding to rows from $C_i$ and columns from $C_j$ and rows from
$C_j$ and columns from $C_i$ are all equal to $\widehat{d}(C_i,C_j)$.
This provides a block decomposition of $D'$ such that in each
rectangle not intersecting the main diagonal of $D'$ all entries are
equal. The rectangles intersecting the main diagonal are squares
defined by the entries located at the intersections of rows and
columns corresponding to a copoint $C_i$. Therefore the recursive call
to $C_i\in \C_p$ corresponds to dealing with the dissimilarity space
$(C_i,d)$ defined by the entries in this diagonal square. The
dissimilarity matrix of $(\C_p,\widehat{d})$ is obtained from $D'$ by
replacing each $|C_i|\times |C_j|$ rectangle by a single entry
$\widehat{d}(C_i,C_j)$ and contracting each copoint of $\C_p$ to a
single point. This illustrates how the dissimilarity space is
decomposed into the dissimilarities of each copoint (the diagonal
blocks) plus the quotient space (representing the non-diagonal
blocks).

\begin{runningexample}
  The copoints of point $1$ give out the block decomposition of
  \Cref{fig:matrix-copoints}.
\end{runningexample}

\begin{figure}[htbp]
  {\footnotesize{
      \begin{equation*}
        \begin{array}{c|c|c|c|c|c|ccccc|c|ccc|ccccc}
             & 1 & 9 & 17 &  6 & 10 &  3 &  4 &  8 & 16 & 18 &  7 & 11 & 13 & 14 &  2 &  5 & 12 & 15 & 19 \\ \hline
          1  & 0 & 4 &  4 &  8 &  8 &  9 &  9 &  9 &  9 &  9 &  9 &  9 &  9 &  9 & 10 & 10 & 10 & 10 & 10 \\ \hline
          9  &   & 0 &  6 &  9 &  8 &  9 &  9 &  9 &  9 &  9 &  9 &  9 &  9 &  9 & 10 & 10 & 10 & 10 & 10 \\ \hline
          17 &   &   &  0 &  8 &  7 &  9 &  9 &  9 &  9 &  9 &  9 &  9 &  9 &  9 & 10 & 10 & 10 & 10 & 10 \\ \hline
          6  &   &   &    &  0 &  7 &  9 &  9 &  9 &  9 &  9 &  9 &  9 &  9 &  9 & 10 & 10 & 10 & 10 & 10 \\ \hline
          10 &   &   &    &    &  0 &  9 &  9 &  9 &  9 &  9 &  9 &  9 &  9 &  9 & 10 & 10 & 10 & 10 & 10 \\ \hline
          3  &   &   &    &    &    &  0 &  1 &  2 &  2 &  2 &  9 &  6 &  6 &  6 &  8 &  8 &  8 &  8 &  8 \\
          4  &   &   &    &    &    &    &  0 &  2 &  3 &  2 &  9 &  6 &  6 &  6 &  8 &  8 &  8 &  8 &  8 \\
          8  &   &   &    &    &    &    &    &  0 &  2 &  2 &  9 &  6 &  6 &  6 &  8 &  8 &  8 &  8 &  8 \\
          16 &   &   &    &    &    &    &    &    &  0 &  2 &  9 &  6 &  6 &  6 &  8 &  8 &  8 &  8 &  8 \\ 
          18 &   &   &    &    &    &    &    &    &    &  0 &  9 &  6 &  6 &  6 &  8 &  8 &  8 &  8 &  8 \\ \hline
          7  &   &   &    &    &    &    &    &    &    &    &  0 &  9 &  9 &  9 & 11 & 11 & 11 & 11 & 11 \\ \hline
          11 &   &   &    &    &    &    &    &    &    &    &    &  0 &  1 &  1 &  5 &  5 &  5 &  5 &  5 \\
          13 &   &   &    &    &    &    &    &    &    &    &    &    &  0 &  1 &  5 &  5 &  5 &  5 &  5 \\
          14 &   &   &    &    &    &    &    &    &    &    &    &    &    &  0 &  5 &  5 &  5 &  5 &  5 \\ \hline
          2  &   &   &    &    &    &    &    &    &    &    &    &    &    &    &  0 &  2 &  2 &  1 &  2 \\
          5  &   &   &    &    &    &    &    &    &    &    &    &    &    &    &    &  0 &  3 &  2 &  1 \\
          12 &   &   &    &    &    &    &    &    &    &    &    &    &    &    &    &    &  0 &  2 &  4 \\
          15 &   &   &    &    &    &    &    &    &    &    &    &    &    &    &    &    &    &  0 &  2 \\
          19 &   &   &    &    &    &    &    &    &    &    &    &    &    &    &    &    &    &    &  0 \\
        \end{array}
        \end{equation*}
    }}
  \caption{A permuted matrix for the copoints attached at $1$.}
  \label{fig:matrix-copoints}
\end{figure}



\subsection{Tree representations of mmodules}\label{sec:mmodules-tree}

A family of subsets $\{M_1,\ldots,M_k\}$ of $X$ is a
\emph{copartition} of $X$ if
$\{X \setminus M_1,\ldots, X \setminus M_k\}$ is a partition of $X$.
For a set $M \subseteq X$, let  $\co{M} := X \setminus M$.
We denote by $\wM:=\wM(X,d)$ the set of all maximal mmodules of $(X,d)$.

\begin{lemma}\label{lemma:partition-copartition}
  $\wM$ is  a partition or a copartition of $X$.
\end{lemma}

\begin{proof}
  If the maximal mmodules are pairwise disjoint, then $\wM$ is a
  partition. We assume now that there exist intersecting maximal
  mmodules $M$ and $M'$. Then $M \cup M' = X$ by
  \Cref{mmodules}\ref{item:mm4}. We assert that every pair of maximal
  mmodules intersects. Let $M_1, M_2 \in \wM$ and suppose $M_1$ and
  $M_2$ are disjoint. Since $M \cup M' = X$, we may assume
  $M \cap M_1 \neq \varnothing$. By \Cref{mmodules}\ref{item:mm4},
  $M \cup M_1 = X$, and then $M_2\subseteq M$. By maximality
  $M_2 = M$, contradicting the fact that $M_1$ and $M_2$ are disjoint.
  Hence any two maximal mmodules $M_1$ and $M_2$ intersect, yielding
  $M_1 \cup M_2 = X$ and $\co{M_1}\cap \co{M_2}=\varnothing$.

  Let $A = \bigcap \wM$ be the intersection of all maximal mmodules,
  and suppose that $A$ is not empty. Then, as $X = \bigcup \wM$, we
  can write $\co{A} = \bigcup_{M_1, M_2 \in \wM} M_1 \setminus M_2$,
  which by \Cref{mmodules}\ref{item:mm3} implies that $\co{A}$ is an
  mmodule. By definition of $A$, $A$ is contained in every
  maximal mmodule, and by assumption $\co{A} \neq X$, thus
  $A \cup \co{A} = X$ is contained in a maximal mmodule,
  contradiction. Thus $A$ is empty. This proves that
  $\co{A} = \bigcup \{\co{M} : M \in \wM\} = X$, and we know that all
  these sets are disjoint, so $\wM$ is a copartition of $X$.
\end{proof}

\Cref{lemma:partition-copartition} describes the structure of maximal
mmodules. To extend that structure to all mmodules, we must understand
how non-maximal modules relates to maximal mmodules. In the case of a
partition, this is settled by \Cref{mmodules}\ref{item:mm5}. The case
of copartitions is the goal of the next Lemma.

\begin{lemma}\label{lemma:mmodules-in-copartition}
  If $\wM= \{M_1,\ldots,M_k\}$ is a copartition, then for any mmodule
  $M\in \M$, either there is $J \subseteq \{1,2,\ldots,k\}$ such that
  $M = \bigcup_{j \in J} \co{M_j}$ or there is
  $i \in \{1,2,\ldots,k\}$ such that $M \subset \co{M_i}$.
\end{lemma}

\begin{proof}
  Let $M$ be an mmodule and suppose that $M$ intersects the
  complements of two maximal mmodules, say
  $M \cap \co{M_1} \neq \varnothing$ and $M \cap \co{M_2} \neq \varnothing$.
  Since $M_1 \cap M \neq \varnothing$, by
  \Cref{mmodules}\ref{item:mm3}, $M_1 \cup M$ is an mmodule which
  strictly contains $M_1$. By maximality of $M_1$, $M_1 \cup M = X$
  and $\co{M_1} \subseteq M$. Consequently,
  $M = \bigcup \{\co{M_i} : M \cap \co{M_i} \neq \emptyset\}$, proving
  the assertion.
\end{proof}

Given a set $X$, a $\cup\cap${\em -tree} on $X$ is a tree
$\mathcal{T}$ with leaf set $X$ and inner nodes labelled by $\cup$ or
$\cap$ which represents a subset $\mathcal{S}(\mathcal{T})$ of the
power set $\mathcal{P}(X)$:
 \begin{itemize}
 \item[(i)] the set of leaves of any node of $\mathcal{T}$ is in
   $\mathcal{S}(\mathcal{T})$,
 \item[(ii)] if a node $N$ is labelled $\cap$, then the set of leaves
   of the union of any proper subset of children of $N$ is in
   $\mathcal{S}(\mathcal{T})$.
  \end{itemize}

\begin{proposition}
  Let $(X,d)$ be a dissimilarity space. There exists a unique
  $\cup\cap$-tree $\mathcal{T}_\mathcal{M}$ on $X$ (up to reordering
  the children of each node) such that
  $\mathcal{M}(X, d) = \mathcal{S}(\mathcal{T}_\mathcal{M})$.
  $\mathcal{T}_\mathcal{M}$ is called the {\em mmodule-tree} of
  $(X, d)$.
\end{proposition}

\begin{proof}
  If $\wM$ is a partition, then the root of $\mathcal{T}_\mathcal{M}$
  has label $\cup$ and its children are the trees defined inductively
  for each maximal mmodule. If $\wM$ is a copartition, then the root
  of $\mathcal{T}_\mathcal{M}$ has label $\cap$ and its children are
  the trees defined inductively for complements of maximal mmodules.
  By \Cref{lemma:partition-copartition}, this procedure defines a
  tree, whence it only remains to establish (i) and (ii). These
  properties hold for maximal mmodules. Pick now a non-maximal mmodule
  $M$. By \Cref{mmodules}\ref{item:mm5}, if $\wM$ is a partition, $M$
  is contained in a maximal mmodule $M'$ associated with some child of
  the root. By induction hypothesis, $M$ is represented in that child.
  If $\wM$ is a copartition, by \Cref{lemma:mmodules-in-copartition},
  either $M$ is the union of the complements of maximal mmodules,
  which corresponds to (ii), or $M$ is strictly contained in the
  complement of some maximal mmodule $M''$, and $\co{M''}$ is
  represented as a child of the root. By induction hypothesis, $M$ is
  represented in that child.
\end{proof}

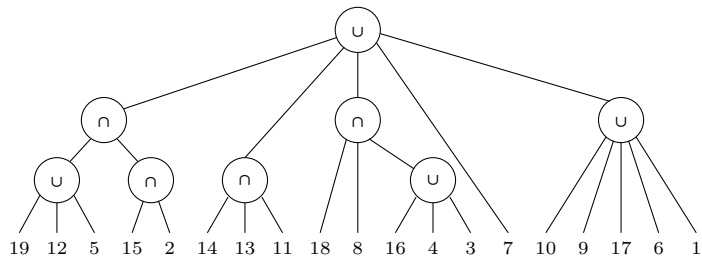
\begin{figure}[htbp]
  \begin{center}
    \begin{tikzpicture}[scale=.5]
      \begin{scope}
        \draw(-1.45, 1) -- (-2, 0) [below] node{\scriptsize{$19$}} ;
        \draw(-1, .8) -- (-1, 0) [below] node{\scriptsize{$12$}} ;
        \draw(-.55, 1) -- (0, 0) [below] node{\scriptsize{$5$}} ;
        \draw(-1, 1.4) circle(.6) ;
        \draw(-1, 1.35) node{\scriptsize{$\cup$}} ;
        \draw(1.3, .85) -- (1, 0) [below] node {\scriptsize{$15$}} ;
        \draw(1.7, .85) -- (2, 0) [below] node{\scriptsize{$2$}} ;
        \draw(1.5, 1.4) circle(.6) ; \draw(1.5, 1.35) node{\scriptsize{$\cap$}} ;
        \draw(-.1, 2.5) -- (-.65, 1.9) ;
        \draw(.6, 2.5) -- (1.15, 1.9) ;
        \draw(0.25, 3) circle(.6) ; \draw(0.25, 2.95) node{\scriptsize{$\cap$}} ;
        \draw(3.55, .95) -- (3, 0) [below] node{\scriptsize{$14$}} ;
        \draw(4, .8) -- (4, 0) [below] node{\scriptsize{$13$}} ;
        \draw(4.45, .95) -- (5, 0) [below] node{\scriptsize{$11$}} ;
        \draw(4, 1.4) circle(.6) ;
        \draw(4, 1.4) node {\scriptsize{$\cap$}};
        \draw(6.7, 2.5) -- (6, 0) [below] node{\scriptsize{$18$}} ;
        \draw(7, 2.4) -- (7,0) [below] node{\scriptsize{$8$}} ;
        \draw(8.55, .95) -- (8, 0) [below] node{\scriptsize{$16$}} ;
        \draw(9, .8) -- (9, 0) [below] node{\scriptsize{$4$}} ;
        \draw(9.45, .95) -- (10, 0) [below] node{\scriptsize{$3$}} ;
        \draw(9, 1.4) circle(.6) ; \draw(9, 1.4) node {\scriptsize{$\cup$}};
        \draw(7.3, 2.5) -- (8.5, 1.7) ;
        \draw(7, 3) circle(.6) ; \draw(7, 2.95) node{\scriptsize{$\cap$}} ;
        \draw(6.64,4.94) -- (4, 2) ; \draw(6.45, 5.2)--(.78, 3.3) ;
        \draw(7, 4.8)--(7, 3.6) ;
        \draw(7, 5.4 ) circle(.6) ; \draw(7, 5.35) node {\scriptsize{$\cup$}};
        \draw(7.5, 5.05) -- (11, 0) [below] node{\scriptsize{$7$}} ;
        \draw(13.6, 2.55) -- (12, 0) [below] node{\scriptsize{$10$}} ;
        \draw(13.8, 2.45) -- (13, 0) [below] node{\scriptsize{$9$}} ;
        \draw(14, 2.4) -- (14, 0) [below] node{\scriptsize{$17$}} ;
        \draw(14.2, 2.45) -- (15, 0) [below] node{\scriptsize{$6$}} ;
        \draw(14.4, 2.55) -- (16, 0) [below] node{\scriptsize{$1$}} ;
        \draw(7.6, 5.3) -- (13.7, 3.5)  ;
        \draw(14, 3) circle(.6) ; \draw(14, 2.95) node{\scriptsize{$\cup$}} ;

      \end{scope}
    \end{tikzpicture}
  \end{center}
  \caption{The mmodule-tree $\mathcal{T}_\mathcal{M}$ of the running example.}
  \label{FIGURE_gros_example}
\end{figure}

\begin{runningexample}
The mmodule-tree for the running example is given in
\Cref{FIGURE_gros_example}. Since the root is a $\cup$-node, the
maximal mmodules are the leafsets of its children.
\end{runningexample}

\section{Flat and conical Robinson spaces}\label{s:flat-conical}

In this section, we first study the copoint partitions in flat Robinson
spaces. They loosely correspond to the Robinson spaces for which our
algorithm find a compatible order without recursion. For conical
Robinson spaces we show how to derive compatible orders from a
compatible order of its subspace obtained by removing the apex. The
importance of conical Robinson spaces stems from the observation that
each copoint of $\C_p$ together with $p$ define a conical subspace
with apex $p$. 

\subsection{Copoint partitions in flat Robinson spaces}

Copoints in flat Robinson spaces (see \Cref{def:flat}) are
characterized by the following result:

\begin{proposition}\label{prop:mmodules-in-unique-order}
  If $(X,d)$ is a flat Robinson space, then either all copoint
  partitions of $(X,d)$ are trivial or there exists a (unique)
  non-diametral point $p$ of $X$ such that $(X,d)$ is conical with
  apex $p$
  and all mmodules of $(X\setminus \{ p\}, d)$
  are trivial.
\end{proposition}

\begin{proof}
  Let $n=|X| > 2$. We order $X$ by a compatible
  order $q_1 < q_2 < \ldots < q_n$. Let $M$ be an mmodule of $(X,d)$.
  Let $i = \min \{ k \in \{1,\ldots,n\} : q_k \in M \}$ and
  $j = \max \{ k \in \{1,\ldots,n\} : q_k \in M \}$. Consider the
  order $<'$ obtained from $<$ by reversing the order between the
  elements in $\{q_i,\ldots,q_j\}$:
  $$q_1 <' q_2 <' \ldots <' q_{i-1} <' q_j <' q_{j-1} <' \ldots <' q_i <' q_{j+1} <' q_{j+2} <' \ldots <' q_n.$$
  We assert that $<'$ is a compatible order. Indeed, let
  $q_x <' q_y <' q_z$, and assume that they are not in the same order
  as in $<$ or $<^\op$. Hence $x < i \leq z < y \leq j$ (or
  symmetrically $i \leq y < x \leq j < z$). Then
  $q_x < q_i \leq q_z < q_y \leq q_j$, from which we get
  $ d(q_x,q_i) \leq d(q_x,q_z) \leq d(q_x,q_y) \leq d(q_x,q_j) = d(q_x,q_i),$
  and then $d(q_y,q_z) \leq d(q_x,q_z) = d(q_x,q_y)$, proving the
  compatibility of $<'$.

  Since $(X,d)$ is flat, $<$
  and $<'$ are either equal or reverse to each other. In the
  first case, this means that $i = j$ and $M$ is trivial. In the
  second case, this means that $i = 1$ and $j = n$.
  So every non trivial mmodule of $(X, d)$  contains $q_1$ and $q_n$.
  Suppose now that
  there are $\alpha < \beta$ in $\{2,\ldots,n-1\}$ with
  $q_\alpha, q_\beta \notin M$. Then
  $d(q_1,q_\alpha) \leq d(q_1,q_\beta) = d(q_\beta,q_n) \leq d(q_\alpha,q_n) = d(q_1,q_\alpha),$
  implying that those quantities are all equal to the same value
  $\delta$. From this, $d(u,v) = \delta$ for each $u \in M$,
  $v \notin M$, hence $X \setminus M$ is also an mmodule. As it does  not
  contain $q_1$ and $q_n$, $X \setminus M$
  is a trivial mmodule, hence $|M| = n - 1$.

  Consequently, any non-trivial mmodule of $(X,d)$ is of the form
  $X \setminus \{q_i\}$ for some $i \in \{2,\ldots,n-1\}$. Suppose
  that $(X,d)$ admits  two non-trivial mmodules
  $X \setminus \{q_i\}$ and $X \setminus \{q_j\}$ with $1 < i < j < n$
  (notice that we need $n \geq 4$). Then for all
  $x \in X \setminus \{q_i,q_j\}$ we have
  $d(x,q_i) = d(q_i,q_j) = d(q_j,x)$, hence $\{q_i,q_j\}$ is an
  mmodule. Since $n > 3$, this is a contradiction to the fact that
  the non-trivial mmodules have cardinality $n-1$. This proves that $M$ is
  unique.

  Finally, let $\Delta = \diam(X) = d(q_1,q_n)$ be the diameter of $(X,d)$,
  let $j$ be such that $M = X \setminus \{q_j\}$.
  Suppose that $q_j$ is the end of a diametral pair, that is
  $d(q_i,q_j) = \Delta$ for any $q_i \in M$. Then for all
  $i \in \{1,\ldots,j-1\}$ and all $k \in \{j+1,\ldots,n\}$, we have
  $\Delta = d(q_i,q_j) \leq d(q_i,q_k) \leq d(q_1,q_n) = \Delta,$
  implying that $\{q_1,\ldots,q_j\}$ is a non-trivial mmodule not
  containing $q_n$, a contradiction.
\end{proof}


\subsection{Conical Robinson spaces}\label{subs:conical} For a
conical Robinson space $(X,d)$ with apex $p$ (see \Cref{def:conical}),
let $d(p,x)=\delta$ for any $x\in X\setminus \{p\}$ and
$X'= X\setminus \{p\}$. We will show how to compute, from any
compatible order $<'$ of $(X',d)$, a compatible order $<$ of $(X,d)$.

Let $<'$ be a compatible order of $(X',d)$. Let $x_*$ and $x^*$ be
respectively the minimal and maximal points of $<'$. By a \emph{hole}
of $<'$ we will mean any pair $(y,z)$ of consecutive points
$y,z \in X'$ of $<'$ with $y <' z$. Informally speaking, a hole is a
place where one can insert the point $p$ and still get a total order.
We will also consider the pair $(x^*,x_*)$ as a hole (this corresponds
to inserting $p$ before or after $X'$). For a hole
$(y,z) \ne (x^*,x_*)$, let $<_{(y,z)}$ be the total order obtained by
inserting $p$ in the hole $(y,z)$, i.e., by setting $u <_{(y,z)} v$
when $u <' v$ if $u, v \in X'$, and $u <_{(y,z)} p$, $p <_{(y,z)} v$
for any $u,v \in X'$ such that $u \le' y$ and $z \le' v$. If
$(y,z)=(x^*,x_*)$, then we set $v<_{(y,z)} p$ for all $v \in X'$ ($p$
is located to the right of $x^*$) or we set $p<_{(y,z)} u$ for all
$u \in X'$ ($p$ is located to the left of $x_*$). We will call a hole
$(y,z)$ of $<'$ \emph{admissible} if $<_{(y,z)}$ is a compatible order
of $(X,d)$.

\begin{lemma}\label{compatible-xy}
  Let $<'$ be a compatible order of $(X',d)$. A hole
  $(y,z) \ne (x^*,x_*)$ of $<'$ is admissible if and only
  if for any $u,v \in X'$ with $u <' v$, the following conditions hold:
  \begin{enumerate}[label=\textup{(\arabic*)},nosep]
  \item\label{item:admhole1} $d(u,v) \ge \delta$ if $u \le' y$ and $z \le' v$;
  \item\label{item:admhole2} $d(u,v) \le \delta$ if $v \le' y$ or $z \le' u$.
  \end{enumerate}
\end{lemma}

\begin{proof}
  Consider a hole $(y,z) \ne (x^*,x_*)$. Pick any three points
  $u, v, w \in X$ such that $u <_{(y,z)} v <_{(y,z)} w$. If
  $p \notin \{u, v, w\}$, then $u <' v <' w$ and thus
  $d(u,w) \ge \max \{d(u,v), d(v,w)\}$. Now, let $p \in \{u,v,w\}$.
  First suppose that $p = u$ (the case $p = w$ is similar). Then
  $d(u,v) = d(u,w) = \delta$. Consequently, $d(v,w) \le d(u,w)$ if and
  only if $d(v,w) \le \delta$, i.e., condition~\ref{item:admhole2}
  holds. Now suppose that $p = v$. Then $d(u,v) = d(v,w) = \delta$. Then
  $d(u,w) \ge \max \{d(u,v), d(v,w)\}$ if and only if
  $d(u,w) \ge \delta$, i.e., condition~\ref{item:admhole1} holds.
  Consequently, the hole $(y,z)$ is admissible if and only if both
  conditions~\ref{item:admhole1} and~\ref{item:admhole2} hold.
\end{proof}

\begin{lemma}\label{lemma:extreme-hole}
  Let $<'$ be a compatible order of $(X',d)$. The hole $(x^*,x_*)$ is
  admissible if and only if $d(x_*,x^*)\le \delta$. Moreover in that
  case, a hole $(y,z) \neq (x^*,x_*)$ is admissible if and only if
  $d(y,z) = \delta$.
\end{lemma}

\begin{proof}
  Since $<'$ is a compatible order of $(X',d)$ and for any
  $u,v \in X'$ we have $x_* \le' u <' v \le' x^*$ or
  $x_* \le' v <' u \le' x^*$, we conclude that
  $d(x_*,x^*) \ge d(u,v)$. Therefore $(x^*,x_*)$ is admissible if and
  only if $\delta \ge d(x_*,x^*)$. The second part of the lemma is a
  consequence of \Cref{compatible-xy}, as
  condition~\ref{item:admhole2} is implied by the fact that the
  diameter is $\delta$.
\end{proof}

\Cref{lemma:extreme-hole} characterizes admissible holes when
$d(x_*, x^*) \leq \delta$. In this case, an admissible hole can be
computed in $O(1)$. The next result provides such a characterization when
$d(x_*, x^*) > \delta$.


\begin{lemma}\label{hole}
  If $(X,d)$ is a conical Robinson space with apex $p$ with
  $\delta < \diam(X')$, and $<'$ is a compatible order on $(X', d)$
  with minimal element $x_*$ and maximal element $x^*$. Then a hole
  $(y,z)$ is admissible if and only if $d(x_*,y) \leq \delta$, $d(z,x^*) \leq \delta$ and
  $d(y,z) \geq \delta$, and such a hole exists.
\end{lemma}

\begin{proof}
  Let $(y,z)$ be a hole with $d(x_*,y) \leq \delta$,
  $d(z,x^*) \leq \delta$ and $d(y,z) \geq \delta$. Then for any
  $x_* \leq' u <' v \leq' y$, $d(u,v) \leq d(x_*,y) = \delta$ holds,
  for any $z \leq' u <' v \leq' x^*$, $d(u,v) \leq d(z,x^*) = \delta$
  holds, and for any $u \leq' y <' z \leq' v$,
  $d(u,v) \geq d(y,z) \geq \delta$ holds. By \Cref{compatible-xy},
  $(y,z)$ is an admissible hole. The reverse implication follows from
  \Cref{compatible-xy}.

  We prove the existence of such a hole. As $x_*$ and $x^*$ are
  extremal, we have $d(x_*, x^*) = \diam(X') = \diam(X) > \delta$.
  Note that the elements of $X'$ are sorted by $<'$ in increasing
  order of their distances from $x_*$, and in decreasing order of
  their distances to $x^*$. Thus the maximal element $a$ such that
  $d(a,x^*) > \delta$ and the minimal element $b$ such that
  $d(x_*,b) > \delta$ (relative to $<'$) are well-defined. Let $<$ be
  a compatible order for $(X,d)$ with $x_* < x^*$. Note that
  $p \in B_\delta(x_*) \cap B_\delta(x^*)$. As $d(x_*,x^*) > \delta$,
  we obtain $x_* < p < x^*$.

  \begin{claim}\label{claim:hole1}
    $X = B_\delta(x_*) \cup B_\delta(x^*)$.
  \end{claim}

  \begin{proof}
    Suppose that there is a point $w \in X$ with
    $\min \{d(w,x_*), d(w,x^*)\} > \delta$. We may assume that
    $w < x_* < p < x^*$. Then $\delta = d(w,p) \geq d(w,x_*) > \delta$,
    a contradiction.
  \end{proof}

  \begin{claim}
    $x_* \leq' a <' b \leq' x^*$.
  \end{claim}

  \begin{proof}
    As $b \notin B_\delta(x_*)$ and by \Cref{claim:hole1},
    $b \in B_\delta(x^*)$, hence $d(b,x^*) < d(a,x^*)$, and thus
    $x_* \leq' a <' b \leq' x^*$.
  \end{proof}

  \begin{claim}
    For any $y \in X$ with $y < p$, $y \in B_\delta(x_*)$ holds. For any
    $y \in X$ with $y > p$, $y \in B_\delta(x^*)$ holds.
  \end{claim}

  \begin{proof}
    If $x_* \leq y < p$, then $d(x_*,y) \leq d(x_*,p) = \delta$ and
    $y \in B_\delta(x_*)$. If $y < x_* < p < x^*$, then
    $d(y,x^*) \geq d(x_*,x^*) > \delta$, implying that
    $y \notin B_\delta(x^*)$, and by \Cref{claim:hole1},
    $y \in B_\delta(x_*)$. The case when $p < y$ is symmetric.
  \end{proof}

  Then $a \notin B_\delta(x_*)$ implies that $a < p$, and similarly
  $p < b$. Thus there exist two consecutive elements $y,z$ for $<'$
  with $a \leq' y <' z \leq' b$ such that $y < p < z$. Then
  $d(y,z) \geq d(y,p) = d(p,z) = \delta$, and $d(x_*,z) \leq \delta$,
  $d(x,x^*) \leq \delta$ by the choice of $a$ and $b$. Thus $(y,z)$ is
  an admissible hole, concluding the proof of the lemma.
\end{proof}

From what precedes, we can derive the following:

\begin{proposition} \label{prop:separateIfSeparableComplexity}
  \Cref{algo:separateIfSeparable} computes an admissible hole and
  separates $X'$ into two subspaces in time $O(|X'|)$.
 \end{proposition}
 In the divide-and-conquer recognition algorithm of Robinson spaces we
 will apply \Cref{algo:separateIfSeparable} with the $p$-copoints of
 $X$ as $X'$.

\begin{algorithm}[htbp]
  \caption{$\textrm{separateIfSeparable}(p,X')$}
  \label{algo:separateIfSeparable}
  \begin{algorithmic}[1]
  \Require{a Robinson space $(X,d)$ (implicit), conical with apex $p$,
    with $X' = X \setminus \{p\}$  sorted along a compatible order $<'$.}
  \Ensure{$X'$ or a bipartition of $X'$, depending on whether $\diam(X')\leq d(p,X')$ or not, each set with a representative.}
  \Let $x_*, x^*$ be the minimum and maximum elements of $(X',<')$
  \Let $\Delta = d(x_*, x^*)$, $\delta = d(p,x_*)$
  \If{ $\Delta \leq \delta$}
      \Return $[ (x_*, X') ]$
  \EndIf
  \For { $y \in X' \setminus \{x^*\}$ in increasing compatible order}
      \Let $z$ be the element consecutive to $x$ in $<'$
      \If{ $d(x_*,y) \leq \delta$ and $d(z,x^*) \leq \delta$ and $d(y,z) \geq \delta$}
        \Return $[ (x_*, \{u \in X' : u \leq' y\}), (x^*, \{ u \in X' : z \leq' u\}) ]$
      \EndIf
  \EndFor
  \end{algorithmic}
\end{algorithm}

\begin{remark}
  \Cref{algo:separateIfSeparable} may not terminate when $X'$
  is not a $p$-copoint of a Robinson space, because the condition tested
  by the second {\bfseries if} statement may never be satisfied. Thus when testing
  whether a dissimilarity space is Robinson, if the space is not
  Robinson, the algorithm may either fail in this procedure, or return
  an order that is not compatible.
\end{remark}

\section{Mmodules in Robinson spaces}\label{s:mmodulesRobinson}

In this section, we investigate the mmodules and the copoint
partitions in Robinson spaces. We classify the copoints of $\C_p$ into
separable, non-separable, and tight. We show that if $(X,d)$ is
Robinson, then there exists a compatible order $<$ in which all
non-separable and tight copoints define intervals of $<$. Furthermore,
we show that separable copoints define two intervals of $<$. Since
each copoint of $\C_p$ together with $p$ define a conical subspace of
$(X,d)$, applying \Cref{algo:separateIfSeparable}, we efficiently
partition each separable copoint into two intervals. This leads to an
extended copoint partition $\C^*_p$ and to the extended quotient space
$(\C^*_p,d^*)$, which is also a Robinson space. Finally, we show how
to derive a compatible order $<$ of $(X,d)$ (satisfying the previous
constraints) from compatible orders of the copoints of $\C_p$ and a
compatible order of the extended quotient $(\C^*_p,d^*)$. While the
compatible orders of copoints are computed recursively, a compatible
order of $(\C^*_p,d^*)$ is computed via proximity orders, introduced
and investigated in \Cref{s:flat}.

\subsection{Copoints in Robinson spaces}\label{separable-tight}

Let $\C_p= \{ C_0=\{ p\}, C_1,\ldots,C_k\}$ be a copoint partition
with attaching point $p$ of a Robinson space $(X,d)$ (see
\Cref{def:copoint-partition}). For a copoint $C_i$, denote by
$\delta_i$ the distance $d(p,x)$ for any point $x\in C_i$ and we
suppose that $i< j$ implies that $\delta_i\le \delta_j$. For
$C_i,C_j\in \C_p, i\ne j$, let $\delta_{ij}$ be the distance $d(x,y)$
between any two points $x\in C_i$ and $y\in C_j$. Since $\C_p$ is a
stable partition, $\delta_{ij}$ is well-defined, moreover
$\delta_{ij}$ coincides with $\widehat{d}(C_i,C_j)$ in the quotient
space $(\C_p,\widehat{d})$. In this subsection, we investigate how in
a compatible order $<$ of $(X,d)$ the copoints of $\C_p$ compares to
the point $p$ and which copoints of $\C_p$ are not comparable to $p$.
Notice that each subspace $(C_i\cup \{ p\},d), i=1,\ldots,k$ is a cone
over $(C_i, d)$ with apex $p$. Applying \Cref{hole} to the Robinson
space $(C_i,d)$ and to the restriction $<_i$ of $<$ to $C_i$, we know
that $<_i$ admits at least one admissible hole. If this hole is
defined by the rightmost and the leftmost points of $<_i$, then $C_i$
is not divided in two parts, otherwise $p$ divides $C_i$ in two parts
$C^l_i$ and $C^r_i$. The next result shows that $C^l_i$ and $C^r_i$
are not only intervals of $<_i$ but also of the global compatible
order $<$:


\begin{lemma}\label{two-parts}
  Let $<$ be a compatible order of $(X,d)$, $p$ be a point of $X$ and
  $C_i \neq \{p\}$ be a copoint of $\C_p$. Then
  $C_i^r := \{u \in C_i : p < u\}$ and
  $C_i^l := \{u \in C_i : u < p \}$ are intervals of $<$ (one of them
  may be empty).
\end{lemma}

\begin{proof}
  Let $C_i^r$ contains at least 2 points (otherwise $C_i^r$ is
  trivially an interval) and let $x,z \in C_i^r$ be its minimum and
  maximum elements, respectively. Let $y \in X$ with $x < y < z$, and
  let $w \in X \setminus C_i$ such that either $w < x$ or $z < w$. If
  $w < x$, then $ d(w,x) \leq d(w,y) \leq d(w,z) = d(w,x),$ where the
  last equality comes from the fact that $C_i$ is an mmodule that
  contains $x,z$ but not $w$. Similarly, if $z < w$, then
  $ d(w,z) \leq d(w,y) \leq d(w,x) = d(w,z).$ Hence for any such $w$,
  the distance $d(w,y)$ is constant for any $y \in [x,z]$. As $C_i$ is
  a maximal mmodule not containing $p$, this implies that
  $[x,z] \subseteq C_i$, that is $C_i^r$ is an interval (possibly
  empty). Symmetrically, $C_i^l$ is also an interval.
\end{proof}

\begin{definition}[Classification of copoints]\label{def:separable}
  A copoint $C_i\in \C_p$ is \emph{separable} if
  $\diam(C_i)>\delta_i$, \emph{non-separable} if
  $\diam(C_i)<\delta_i$, and \emph{tight} if $\diam(C_i)=\delta_i$.
\end{definition}

\begin{runningexample}
  In the running example, the copoints of $\C_1$ are non-separable.
  The copoint $C_3 = \{5,12,19\}$ of $\C_2$ is separable and its two
  halved copoints are $C'_3=\{5, 19\}$ and $C''_3= \{12\}$. 
\end{runningexample}

\begin{lemma}\label{separable}
  If $<$ is a compatible order of $(X,d)$, then any separable copoint
  $C_i$ defines two intervals $C^l_i$ and $C^r_i$ of $<$ such that
  $d(x,y) \ge \delta_i$ for any $x \in C^l_i, y \in C^r_i$ and
  $\diam(C^l_i) \le \delta_i,$ $\diam(C^r_i) \le \delta_i$.
\end{lemma}

\begin{proof}
  Because $\{p\} \cup C_i$ is conical with apex $p$, by \Cref{hole},
  in the restriction of $<$ to $C_i$, the hole $(y,z)$ with
  $y < p < z$ is admissible. Then the two intervals from
  \Cref{two-parts} are $C_i^l := \{x \in C_i~:~x \leq y\}$ and
  $C_i^r := \{ x \in C_i~:~x \geq z\}$, and the result follows by
  \Cref{compatible-xy}
\end{proof}


\begin{lemma}\label{nonseparable}
  If $<$ is a compatible order of $(X,d)$, then any non-separable
  copoint $C_i$ defines a single interval of $<$.
\end{lemma}

\begin{proof}
  If $x<p<y$ for $x,y\in C_i$, then
  $d(x,y)\ge \max\{ d(x,p),d(y,p)\}=\delta_i$, which is impossible
  because $d(x,y)\le \diam(C_i)<\delta_i$. Thus $C_i$ must be located
  on one side of $p$. By \Cref{two-parts}, $C_i$ is an interval
  of $<$.
\end{proof}

\begin{lemma}\label{tight}
  Let $C_i\in \mathcal{C}_p$ be a tight copoint of $(X,d)$ and $<$ be
  a compatible order such that the intervals
  $C_i^r := \{u \in C_i : p < u\}$ and
  $C_i^l := \{u \in C_i : u < p \}$ are not empty. Then the order $<'$
  defined by the rule:
  $$ \textrm{for any $u < v$, if $C_i^\ell < u < C_i^r$ and $v \in C_i^r$, then $v <' u$, otherwise $u <' v$,}
  $$
  is a compatible order of $(X,d)$. Consequently, if $(X,d)$ is
  Robinson, then there exists a compatible order for which each tight
  copoint is a single interval.
\end{lemma}

The order $<'$ is thus obtained from $<$ by moving $C_i^r$ immediately
after $C_i^\ell$. By symmetry, one could get a similar result by moving instead
$C_i^\ell$ in front of $C_i^r$.

\begin{proof}
  Pick any three points $x,y,z \in X$ such that they are not
  identically ordered by $<$ and by $<'$. Then we can suppose without
  loss of generality that $x \in C_i^r$ and $C_i^\ell < y < C_i^r$.
  Notice also that $C^\ell_i\cup C^r_i$ is an interval of $<'$. We
  will show now that whatever is the order of $x,y,z$ with respect to
  $<$, it does contradiction the compatibility of $<'$. First, let
  $z \in X\setminus C_i$. Let $x_\ell$ be any point of $C^\ell_i$.
  Since $C^\ell_i\cup C^r_i$ is an interval of $<'$ containing
  $x,x_\ell$ and not containing $y,z$, the order of $x, y, z$ along
  $<'$ is the same as the order of $x_\ell, y, z$ along $<'$, which is
  the same as the order of $x_\ell, y, z$ along $<$. Since
  $d(x,z)=d(x_\ell,z),d(x,y)=d(x_\ell, y)$ and $<$ is a compatible
  order, the result follows.

  Now, let $z\in C^r_i$. Then we can suppose that $y < x < z$ (the
  other case $y < z <x$ is similar) and thus
  $d(y,z) \geq \max\{d(y,x), d(x,z)\}$. In this case, we have
  $x<'z<'y$. As $d(y,z)=d(y,x)$, we obtain
  $d(y,x) \geq \max\{d(y,z), d(x,z)\}$ and we are done. Finally, let
  $z \in C^\ell_i$. Then we have $z < y < x$. If $p <y$ or $p=y$ (the
  case $y<p$ is symmetric), then since $C_i$ is a tight copoint, we
  obtain
  $\delta_i = d(z,p) \leq d(z,y) = d(y,x) \leq d(z,x) = \delta_i$.
  Consequently, $d(y,x) = d(y,z) = d(x,z)$ and thus the triple $x,y,z$
  yields no contradiction for $<'$.
\end{proof}

\Cref{separable,nonseparable,tight} imply the following result:

\begin{proposition}\label{location-copoints}
  If $(X,d)$ is a Robinson space and $\C_p$ is a copoint partition of
  $X$, then there exists a compatible order $<$ in which each copoint
  $C_i$ with $\diam(C_i)\le \delta_i$ is an interval of $<$ located
  either to the left or to the right of $p$ and each copoint $C_i$
  with $\diam(C_i)>\delta_i$ defines two intervals $C^l_i$ and $C^r_i$
  of $<$ such that $C^l_i<p<C^r_i$.
\end{proposition}

Next we consider only compatible orders of $(X,d)$
satisfying the conditions of \Cref{location-copoints}.

\subsection{Compatible orders from compatible orders of copoints and extended quotient}

Let $\C_p=\{ C_0=\{ p\}, C_1,\ldots,C_k\}$ be a copoint partition with
attaching point $p$ of a Robinson space $(X,d)$. For a separable
copoint $C_i$ an \emph{admissible bipartition} is a partition $C_i$
into $C'_i$ and $C''_i$ such that
$\diam(C'_i)\le \delta_i, \diam(C''_i)\le \delta_i$, and
$d(x,y)\ge \delta_i$ 
for any $x\in C'_i$ and $y\in C''_i$. This
partition is defined by applying \Cref{hole} to each
$(C_i,d),i=1,\ldots,k$ and to each conical Robinson space
$(C_i\cup \{ p\}, d)$. Notice that $C'_i$ and
$C''_i$ are no longer mmodules because the distances between the
points of $C'_i$ and $C''_i$ are not necessarily the same. We will
call $C'_i$ and $C''_i$ \emph{halved copoints}.

\begin{definition} [Extended quotient]
  Let $\C_p^*$ denote the set of all non-separable and tight copoints
  of $\C_p$ plus the set of all halved copoints corresponding to a
  choice of admissible bipartitions of each separable copoint. An
  \emph{extended quotient} of $(X,d)$ is the dissimilarity space
  $(\C^*_p,d^*)$ defined in the following way: for $i \ne j$, the
  distance $d^*(\alpha,\beta)$ between a pair of copoints or halved
  copoints $\alpha,\beta$, (1) is $\delta_{ij}$ when one is indexed by
  $i$ and the other is indexed by $j$, and (2) is $\diam(C_i)$, the
  diameter of $C_i$, when $\alpha$ and $\beta$ are the two half
  copoints from the same copoint $C_i$. Notice that for any points
  $u\in \alpha$ and $v\in \beta$, in the first case we have
  $d^*(\alpha,\beta)=d(u,v)$ and in the second case, we have
  $d^*(\alpha,\beta)\ge d(u,v)$.
\end{definition}

\begin{runningexample}
  The extended quotient space $(\mathcal{C}^*_2,d^*)$ of the running
  example is given in \Cref{fig:ext-quotient-space}. It is unique as
  $C_2$ has a unique admissible bipartition.

\begin{figure}
  {\footnotesize{
      \begin{equation*}
        \begin{array}{cccccccccc}
          && C_0 & C_1 & C_2 & C'_3 & C''_3 & C_4 & C_5 & C_6 \medskip\\
          C_0 && 0  & 10  &  8  &  2 & 2  &  11 &  5 &   1  \\
          C_1 &&    &  0  &  9  & 10 & 10  & 9   & 9  &  10  \\
          C_2 &&    &     &  0  &  8 & 8  &  9  &  6 &  8   \\
          C'_3 &&    &     &     &   0 & 3 &  11 & 5  & 2    \\
          C''_3 &&    &     &     &   & 0 &  11 & 5  & 2    \\
          C_4 &&    &     &     &     & &   0 &  9 &  11  \\
          C_5 &&    &     &     &     &  &   &  0 &   5  \\
          C_6 &&    &     &     &     &   &  &    &   0  \\
        \end{array}
      \end{equation*}
    }}
  \caption{The extended quotient space $(\mathcal{C}^*_2,d^*)$.}
  \label{fig:ext-quotient-space}
\end{figure}
\end{runningexample}

\begin{lemma}\label{extquotient}
  If $(X,d)$ is a Robinson space, then for any $p \in X$, its quotient
  $(\C_p,\widehat{d})$ (see \Cref{def:quotient}) and its extended
  quotient $(\C^*_p,d^*)$ are Robinson spaces.
\end{lemma}

\begin{proof}
  It suffices to isometrically embed $(\C_p,\widehat{d})$ in
  $(\C^*_p,d^*)$ and $(\C^*_p,d^*)$ in $(X,d)$. The map
  $\widehat{\varphi}: \C_p\rightarrow \C^*_p$ maps any non-separable
  or tight copoint $C_i$ to itself and any separable copoint $C_i$ to
  one of its halves. From the definitions of $(\C_p,\widehat{d})$ and
  $(\C^*_p,d^*)$, $\widehat{\varphi}$ is an isometric embedding. The
  map $\varphi^*: \C^*_p\rightarrow X$ is defined as follows. We
  select one point $x_i$ in each non-separable or tight copoint $C_i$
  and set $\varphi^*(C_i)=x_i$ and we select a diametral pair
  $\{ x'_i,x''_i\}$ for each separable copoint $C_i$ separated into
  $C'_i$ and $C''_i$ and set
  $\varphi^*(C'_i)=x'_i, \varphi^*(C''_i)=x''_i$. From the definition
  of $(\C^*_p,d^*)$, $\varphi^*$ is an isometric embedding.
\end{proof}

Now, we will show that if $(X,d)$ is a Robinson space, then from any
compatible order $<^*$ of an extended quotient $(\C^*_p,d^*)$ and
from the compatible orders $<_i$ of the copoints $(C_i, d)$ of $\C_p$,
we can define a compatible order $<$ of $(X,d)$. We recall that any
compatible order $<_i$ of a separable copoint $C_i$ has an
admissible bipartition $\{ C'_i,C''_i\}$ of $C_i$ as defined in
\Cref{subs:conical}. The total order $<$ is defined as follows: for
two points $x,y$ of $X$ we set $x<y$ if and only if (1)
$x\in \alpha, y\in \beta$ for two different points
$\alpha,\beta\in \C^*_p$ and $\alpha <^* \beta$ or (2)
$x,y\in \alpha\in \C^*_p$, $\alpha\subseteq C_i$, and $x<_i y$.

\begin{proposition}\label{quotient+copoints}
  If $(X,d)$ is a Robinson space, then $<$ is a compatible order of
  $(X,d)$.
\end{proposition}

\begin{proof}
  Pick any three distinct points $u<v<w$ of $X$ and let
  $u\in \alpha, v\in \beta$, and $w\in \gamma$ with
  $\alpha,\beta,\gamma\in \C^*_p$. If $\alpha=\beta=\gamma\subseteq C_i$,
  then $u<_i v<_i w$ and the result follows from the fact that $<_i$ is
    a compatible order of $(C_i, d)$. Now, let $\alpha, \beta$, and $\gamma$ be distinct.
Then $\alpha<^* \beta<^* \gamma$ and
    $d^*(\alpha,\gamma) \ge \max\{ d^*(\alpha,\beta),d^*(\beta,\gamma)\}$.
    Since $d^*(\alpha,\gamma) \ge d(u,w)$, $d^*(\alpha,\beta)\ge d(u,v)$, and
    $d^*(\beta,\gamma)\ge d(v,w)$, the inequality
    $d(u,w) \ge \max \{ d(u,v),d(v,w)\}$ holds if
    $d^*(\alpha,\gamma)=d(u,w)$. Now suppose that say
    $d^*(\alpha,\beta)>d(u,v)$. From the definition of $d^*$ this implies
    that $\alpha$ and $\gamma$ are the halved copoints $C'_i$ and $C''_i$
    of $C_i$ and say $C'_i<^*\beta\leq^*p<^*C''_i$. Then $\beta$ belongs to
    a copoint $C_j$ with $\delta_j\le \delta_i$. Since $u,w\in C_i$ and
    $v\in C_j$,
    $d(u,v)=d(v,w)=\delta_{ij}=d^*(\beta,\gamma)=d^*(\alpha,\beta)\le \delta_i$.
    On the other hand, since $\{ C^l_i,C^r_i\}$ is an admissible partition
    of $C_i$, $d(u,w)\ge \delta_i$ and we are done.

Finally, let $\alpha=\beta$ or $\beta=\gamma$, say the first.
First suppose that $\alpha$ and $\gamma$ are halved copoints of
    $C_i$: $\alpha = C^l_i$ and $\gamma = C^r_i$. Then $u <_i v <_i w$
    and since $<_i$ is a compatible order of $(C_i, d)$, we are
    done. Now suppose that $\alpha$ and $\gamma$ belong to different
    copoints, say $\alpha \subseteq C_i$ and $\gamma \subseteq C_j$.
    Since $u,v \in C_i$ and $w \in C_j$,
    $d(u,w) = d(v,w) = \delta_{ij}$. It remains to prove that
    $d(u,v) \le \delta_{ij}$. By \Cref{location-copoints},
    there is a compatible order for which $\alpha$ is an interval, in
    particular for which $w$ is not between $u$ and $v$. This implies
    $d(u,v) \leq \max\{d(u,w), d(v,w)\} = \delta_{ij}$.
    This concludes the proof of the proposition.
\end{proof}

\section{Proximity orders}\label{s:flat}

In this section, we introduce the concepts of $p$-proximity order and
$p$-proximity pre-order for a compatible order of $(X,d)$. We show
that $p$-proximity pre-orders can be efficiently computed by computing
and ordering the copoints of $p$. For $p$-trivial Robinson spaces (see
\Cref{def:trivial-cotrivial}), in particular for quotient spaces, the
$p$-proximity pre-orders are $p$-proximity orders. Furthermore, we
prove that extended quotients of Robinson spaces (even if they are not
$p$-trivial) still admit $p$-proximity orders, which can be derived
from $p$-proximity pre-orders. In all those algorithmic results, we
compute a $p$-proximity pre-order or a $p$-proximity order without the
knowledge of any compatible order. The main result of this section is
the description of an efficient algorithm showing, given a
$p$-proximity order $\prec$, how to compute a compatible order for
which $\prec$ is a $p$-proximity order. Applied to the extended
quotients, this algorithm will be the merging step of our
divide-and-conquer recognition algorithm described in
\Cref{s:divide-and-conquer-copoint}. Consequently, the recursive calls
in our recognition algorithm will be applied only to the copoints of
$p$ and not to the extended quotient.

\subsection{\texorpdfstring{$p$}{p}-Proximity orders}
\label{section:pProximityOrder}
We start with the definition of a $p$-proximity order.

\begin{definition}[$p$-Proximity orders and pre-orders]
  Let $(X,d)$ be a Robinson space  with a compatible order
  $<$ and let $p$ be a point of $X$. A \emph{$p$-proximity order}
  for $<$ is a total order $\prec$ on $X$ such that
  \begin{enumerate}[label=(PO\arabic*),nosep]
  \item\label{item:proxorder1} for all distinct $x, y \in X$, if
    $x \prec y$, then $d(p,x) \leq d(p,y)$,
  \item\label{item:proxorder2} for all distinct
    $x,y \in X \setminus \{p\}$, $x \prec y$ implies that either
    $y < p$ and $y < x$, or $y > p$ and $y > x$ ({\it i.e.} $y$ is not
    between $p$ and $x$ in the compatible order $<$). Equivalently,
    for all $x\in X\setminus\{p\}$, the set $\{t\in X: t\prec x\}$ is
    an interval for $<$.
  \end{enumerate}
  A \emph{$p$-proximity pre-order} for a compatible order $<$ is a
  pre-order $\preceq$ on $X$ which can be refined
  to a $p$-proximity order for $<$, and two distinct elements are
  equal only if they belong to the same copoint of $\mathcal{C}_p$.
\end{definition}

Notice that $p$ is the minimum of $\prec$ and that if
$d(p,x) < d(p,y)$, then $y < p$ and $y < x$ or $y > p$ and $y > x$.
Intuitively, given a compatible order $<$, a $p$-proximity order
$\prec$ for $<$ can be obtained by shuffling the elements smaller than
$p$ in reverse order into the elements larger than $p$, and adding $p$
as the minimum.

For two disjoint sets $S,S'$ of $X$, we denote $S \prec S'$ when
for each $x \in S$, $y \in S'$, either $y < p$ and $y < x$ or $y > p$
and $y > x$. We denote $S \leq_p S'$ when for each $x \in S$,
$y \in S'$, $d(p,S) \leq d(p, S')$.

One can build a $p$-proximity pre-order for a compatible order $<$
without the knowledge of $<$. We will use \Cref{algo:recursiveRefine},
which is a variant of \Cref{algo:partitionRefine}
($\variable{partitionRefine}$) and also uses \Cref{algo:refine}
($\variable{refine}$). It differs from the stable partition algorithm
in making a distinction between in-pivots ($\variable{In}$) and
out-pivots ($\variable{Out}$), that are respectively smaller and
bigger in the $p$-proximity order than the set $S$ to refine. When
using an out-pivot, the output of $\variable{refine}$ must be
reordered, on Line~\ref{step:reverse-prefix}. Notice that when $m=1$,
that is when the call to $\variable{refine}$ on Line~\ref{step:def-m}
returns a single part, the algorithm calls itself on
Line~\ref{step:recursiveRefine:recursive-call} with the same
parameters except that $q$ is removed from the sets of pivots.

\begin{algorithm}[htbp]
  \caption{$\textrm{recursiveRefine}(p,\variable{In},S,\variable{Out})$}
  \label{algo:recursiveRefine}
  \begin{algorithmic}[1]
  \Require{a Robinson space $(X,d)$ (implicit), a point $p \in X$, a set $S \subseteq X$,
    two disjoint subsets $\variable{In}, \variable{Out} \subseteq X \setminus S$ of inner-pivots and outer-pivots.
  }
  \Ensure{an ordered partition $[S^\star_1,S^\star_2,\ldots,S^\star_{k^\star}]$ of $S\setminus\{p\}$ (encoding a partial $p$-proximity pre-order).}
  \If{$\variable{In} \cup \variable{Out} = \emptyset$}
      \Return $[ S ]$\label{step:rrreturn}
  \EndIf
  \Let $q \in \variable{In} \cup \variable{Out}$ \Comment{choose $q$ to be the first element of $\variable{In}$ or $\variable{Out}$}\label{recRefine_def-q}
  \Let $[S_1,\ldots,S_m] = \textrm{refine}(q,S)$ \label{step:def-m}
  \If {$q \in \variable{Out}$}
      \Let $\alpha = \min( \{ j \in \{1,\ldots,m\} : d(S_j,q) > d(p,q)\} \cup \{m+1\})$ \label{step:def-alpha}
      \Let $[S'_1,\ldots,S'_m] = [S_{\alpha-1}, S_{\alpha-2}, \ldots, S_1, S_\alpha, S_{\alpha+1}, \ldots, S_m]$ \label{step:reverse-prefix}
  \Else
      \Let $[S'_1, \ldots,S'_m] = [S_1,\ldots,S_m]$ \label{step:recursiveRefine:beforeRecursiveCalls}
  \EndIf
  \For {$i \in \{1,\ldots,m\}$}
      \Let $\variable{In}_i = \textrm{concatenate}(S'_1, \ldots,S'_{i-1}, \variable{In} \setminus \{q\})$ \label{step:def-ini}
      \Let $\variable{Out}_i = \textrm{concatenate}(S'_{i+1}, \ldots, S'_m, \variable{Out} \setminus \{q\})$ \label{step:def-outi}
      \Let $T_i = \textrm{recursiveRefine}(p,\variable{In}_i,S'_i,\variable{Out}_i)$ \label{step:recursiveRefine:recursive-call}
  \EndFor
  \Return $\textrm{concatenate}(T_1,\ldots,T_m)$\label{step:concat-all}
  \end{algorithmic}
\end{algorithm}

\begin{lemma}\label{lemma:rrconditions}
  Let $(X,d)$ be a Robinson space. Let $\variable{In}$, $S$ and
  $\variable{Out}$ be disjoint subsets of $X$, and
  $p \in X \setminus S$. Suppose that:
  \begin{enumerate}[label=(\roman*),ref=(\roman*)]
  \item\label{item:rr1} $\variable{In} \leq_p S \leq_p\variable{Out}$,
  \item\label{item:rr2} $\variable{In} \prec S \prec \variable{Out}$,
  \item\label{item:rr3} for all $x,x' \in S$ and
    $y \in X \setminus (\variable{In} \cup S \cup \variable{ Out})$,
    $d(x,y) = d(x',y)$,
  \end{enumerate}
  and either $(\variable{In},\variable{Out}) = (\{p\},\varnothing)$ or
  $p \notin \variable{In} \cup S \cup \variable{Out}$. Let
  $[S_1^\star,\ldots,S_{k^\star}^\star]$ be the output of
  \Cref{algo:recursiveRefine} with input
  $(p,\variable{In},S,\variable{Out})$.  Then the following properties hold: 
  \begin{enumerate}[resume, label=(\roman*), ref=(\roman*)]
  \item\label{item:rr4} $\{S_1^\star,\ldots,S_{k^\star}^\star\}$ is a partition of $S$
    into mmodules,
  \item\label{item:rr5} $S_1^\star \prec S_2^\star \prec \ldots \prec S_{k^\star}^\star$.
  \end{enumerate}
\end{lemma}

\begin{proof}
  The proof is by induction on the call tree. When
  \Cref{algo:recursiveRefine} returns at Line~\ref{step:rrreturn},
  \ref{item:rr4} follows from~\ref{item:rr3} ($S$ is an mmodule),
  and \ref{item:rr5} is trivial. We assume now that it returns at
  Line~\ref{step:concat-all} and we will use the notation of the
  algorithm. We prove that the
  conditions~\ref{item:rr1},~\ref{item:rr2},~\ref{item:rr3} holds for
  each recursive call at
  Line~\ref{step:recursiveRefine:recursive-call}. Let
  $i \in \{1,\ldots,m\}$. Notice that by construction
  $\variable{In}_i$, $S'_i$ and $\variable{Out}_i$ are disjoint and
  $p \notin \variable{In}_i \cup S'_i \cup \variable{Out}_i$.

  If $\variable{In} = \{p\}$ and $\variable{Out} = \varnothing$, then
  by \Cref{lemma:refine-analysis} applied to the call to
  $\variable{refine}$, $S'_1 \leq_p S'_2 \leq \ldots \leq S'_m$, from
  which~\ref{item:rr1} follows for $i$. Moreover in that case, for
  any $x \in S'_j$ and $y \in S'_{j'}$ with $j < j'$,
  $d(p,x) < d(p,y)$ implies that either $y < p$ and $y < x$ or $y > p$ and
  $y > x$, thus $S'_j \prec S'_{j'}$, proving~\ref{item:rr2} for each
  $i$.

  Otherwise, $p \notin \variable{In} \cup S \cup \variable{Out}$, thus
  by~\ref{item:rr3}, $d(p,x) = d(p,x')$ for any $x, x' \in S$, also
  proving $\ref{item:rr1}$ for $i$. Condition~\ref{item:rr2} follows
  from the next claim.

  \begin{claim}\label{claim:rr}
    If $p \neq q$, then for each $i < j$, $S'_i \prec S'_j$.
  \end{claim}

  \begin{proof}
    We may assume that $q < p$ (the case $p<q$ is analogous). 
    If $q \in \variable{In}$, then for any $y \in S$ we have
    $y < q < p < y$ or $q < p < y$ by \ref{item:rr2}. Let $x \in S'_i$ and
    $y \in S'_j$.  By \Cref{lemma:refine-analysis}, $d(q,x) < d(q,y)$.
    Hence we have either $y<x<q<p$, or $y<q<p<x$, or $x<q<p<y$, or
    $q<p<x<y$. In any of these cases, $y$ is not between $p$ and $x$,
    hence $S'_i \prec S'_j$.

    If $q \in \variable{Out}$, then for any $y \in S$, we have
    $q < y < p$ or $q < p < y$ by \ref{item:rr2}. Moreover if
    $d(y,q) > d(p,q)$ then $q < p < y$, and for any $x \in S$ with
    $d(x,q) < d(y,q)$, $q < x < y$. Let $x \in S'_i$ and $y \in S'_j$,
    with $i < j$. If $i < \alpha \leq j$, then
    $d(q,x) \leq d(q,p) < d(q,y)$ and $q < x < p < y$ or
    $q < p < x < y$. If $i < j < \alpha$ then
    $d(q,y) < d(q,x) \leq d(q,p)$ and $q < y < x < p$ or
    $q < y < p < x$. If $\alpha \leq i < j$,  then
    $d(q,p) < d(q,x) < d(q,y)$ and $q < p < x < y$. In every case,
    $y$ is not between $p$ and $x$, hence $S'_i \prec S'_j$.
  \end{proof}

  Let $x,x' \in S'_i$. By~\ref{item:rr3} on the input, for each
  $y \in X \setminus (\variable{In} \cup S \cup \variable{Out})$, we
  have $d(x,y) = d(x',y)$. Moreover, by \Cref{lemma:refine-analysis},
  for all $x, x' \in S'_i$, we have $d(q,x) = d(q,x')$. Hence~\ref{item:rr3}
  for $i$ follows by observing that
  $X \setminus (\variable{In}_i, S'_i, \variable{Out}_i) = X \setminus (\variable{In} \cup S \cup \variable{Out}_i) \cup \{q\}$.

  Having proved all the hypothesis for each recursive call, we get that
  for each $i$, \ref{item:rr4} and~\ref{item:rr5} hold: $T_i$ is a
  partition of $S'_i$ into mmodules sorted by $\prec$. From
  Line~\ref{step:concat-all} and because $\{S'_1,\ldots S'_m\}$ is a
  partition of $S$, we obtain that~\ref{item:rr4} holds. Finally, for
  any $S^\star_i$ and $S^\star_j$ with $i < j$, either there is $k$
  with $S^\star_i, S^\star_j \subset S'_k$, in which case
  by~\ref{item:rr5} for recursive call $k$,
  $S^\star_i \prec S^\star_j$, or there are $k < k'$ with
  $S^\star_i \in S'_k$ and $S^\star_j \in S'_{k'}$, in which case
  $S'_k \prec S'_{k'}$ by \Cref{claim:rr}, hence
  $S^\star_i \prec S^\star_j$, proving~\ref{item:rr5}.
  \end{proof}

\begin{proposition}\label{prop:proximity-preorder}
  Let $(X,d)$ be a Robinson space and let $p$ be any point of $X$.
  Then \Cref{algo:recursiveRefine} with input $(p,$
  $\{p\}, X\setminus\{p\}, \varnothing)$ returns the copoint partition
  $\mathcal{C}_p=(C_0=\{ p\}, C_1,\ldots,C_k\})$ of $p$ sorted along a
  $p$-proximity pre-order $\prec$ for any (unknown) compatible order
  $<$ of $(X,d)$.
\end{proposition}

\begin{proof}
  We can apply \Cref{lemma:rrconditions} on the initial call to
  \Cref{algo:recursiveRefine}, from which it only remains to prove
  that each mmodule $S^\star_i$ is a $p$-copoint. Suppose it is not a
  copoint, then it is contained in a copoint $C \subset S$. Consider
  the deepest recursive call of \Cref{algo:recursiveRefine} for which
  $C \subseteq S$. Then it must be that on Line~\ref{step:def-m},
  \variable{refine} splits $C$ into several parts, which means that
  there is $x,y \in C$ such that $d(q,x) \neq d(q,y)$ by
  \Cref{lemma:refine-analysis}, for $q \notin C$. But this contradicts
  the fact that $C$ is an mmodule.
\end{proof}

Now, we analyse the complexity of \Cref{algo:recursiveRefine}
without counting the time spent by all the recursive calls to
$\mathit{refine}$. This will be done at a later stage in our
analysis.

\begin{lemma}\label{lemma:recursiveRefine-complexity}
  Without counting the time spent in the calls to \Cref{algo:refine},
  \Cref{algo:recursiveRefine} with input $(p,[q],S,[])$ and output
  $[S^\star_1,\ldots,S^\star_{k^\star}]$ runs in time
  $O(|S|^2 - \sum_{j = 1}^{k^\star} |S^\star_j|^2)$.
\end{lemma}

\begin{proof}
  We consider the tree of recursive calls, where each node correspond
  to some subset of $S$, and the leaves correspond to
  $[S^\star_1,S^\star_2,\ldots,S^\star_{k^\star}]$. Firstly we bound
  the number of nodes in that tree. To this end, notice that for each
  leaf $S^\star_j$, for $j' \neq j$, each element $x \in S^\star_{j'}$
  has been added to either $\variable{In}_i$ or $\variable{Out}_i$ at
  the recursive call corresponding to the lowest common ancestor of
  $S^\star_j$ and $S^\star_{j'}$. Because each recursive call removes
  exactly one element from $\variable{In}_i \cup \variable{Out}_i$,
  the depth of $S^\star_j$ is $|S \setminus S^\star_j|$. Thus the
  number of nodes is at most
  $C := \sum_{j=1}^{k^\star} |S^\star_j| |S \setminus S^\star_j|
  = |S|^2 - \sum_{j=1}^{k^\star} |S^\star_k|^2$.

  Then, the sum of all $m$ (defined on line~\ref{step:def-m}) over all
  recursive calls is the sum of arities of the nodes of the tree,
  hence is less than $C$. This implies that the operations on
  lines~\ref{step:def-alpha}, \ref{step:reverse-prefix}
  and~\ref{step:concat-all}, as well as the cost of constant-time
  operations of each call, contribute $O(C)$ to the total cost of the
  algorithm. It remains to bound the cost of lines~\ref{step:def-ini}
  and~\ref{step:def-outi}. But the cost of inserting elements into
  $\variable{In}_i$ and $\variable{Out}_i$ can be no more that the
  cost of removing all those inserted elements, which we have already
  bounded by $C$.
\end{proof}

For $p$-trivial Robinson spaces, the $p$-proximity pre-orders are orders:

\begin{proposition}\label{prop:proximity-order}
  If $(X,d)$ is a $p$-trivial Robinson space, then any $p$-proximity
  pre-order is a total order and it can be computed in $O(|X|^2)$
  time.
\end{proposition}

\begin{proof}
  Since all copoints of $\C_p$ are trivial, a $p$-proximity pre-order
  $\preceq$ is by definition a total order, hence $\preceq$ is a
  $p$-proximity order. By \Cref{prop:proximity-preorder}, this
  $p$-proximity order can be computed in $O(|X|^2-|X|+1)$ time because
  $k=|X|-1$ and all copoints have size 1.
\end{proof}

\begin{runningexample}
  First consider \Cref{algo:recursiveRefine} for $p=1$ on input
  $(p,\{p\},X \setminus \{p\}, \varnothing)$. Then on
  Line~\ref{step:def-m}, using $q = p = 1$, we get an ordered
  partition
  $$[\{9,17\}, \{6,10\}, \{3,4,7,8,11,13,14,16,18\}, \{2,5,12,15,19\}].$$
  Since $q \in \variable{In}$, the order is not modified and the
  recursive calls output respectively $[\{17\},\{9\}]$,
  $[\{10\},\{6\}]$, $[\{3,4,8,16,18\},\{11,13,14\},\{7\}]$ and
  $[\{2,5,12,15,19\}]$. Thus the ordered copoint partition is:
  $$ [\{1\}, \{17\}, \{9\}, \{10\}, \{6\}, \{3,4,8,16,18\}, \{11,13,14\}, \{7\}, \{2,5,12,15,19\}].$$

  To go deeper, let us detail the third recursive call, with parameters
  $$(p, \variable{In} = \{9,17,6,10\}, S = \{3,4,7,8,11,13,14,16,18\}, \variable{Out} = \{2,5,12,15,19\}\}).$$
  For any element in $\variable{In}$, no refinement happens ($m=1$),
  hence there are multiple recursive calls until getting to parameters
  $(p, \variable{In} = \{\}, S = \{3,4,7,8,11,13,14,16,18\}, \variable{Out} = \{2,5,12,15,19\}\})$.
  Then when choosing $q = 2$ from $\variable{Out}$, the call to
  $\variable{refine}$ on Line~\ref{step:def-m} returns a non-trivial
  ordered partition $[\{11,13\}, \{3,4,8,16,18\}, \{7\}]$, where
  $d(2,11) = 5$, $d(2,3) = 8$, $d(2,7) = 11$ and $d(p,2) = 10$. Then
  after Line~\ref{step:reverse-prefix}, the order is changed to
  $[\{3,4,8,16,18\}, \{11,13\}, \{7\}]$. As each of these parts is a
  copoint, they will not get subdivided further in subsequent
  recursive calls, and this ordered partition is returned.
\end{runningexample}

\subsection{\texorpdfstring{$p$}{p}-Proximity orders for extended quotients}

Let $(X,d)$ be a Robinson space and $p\in X$. By \Cref{rcd*},
$(\C_p,\widehat{d})$ is $p$-trivial. By \Cref{extquotient},
$(\C_p,\widehat{d})$ and $(\C^*_p,d^*)$ are Robinson, but
$(\C^*_p,d^*)$ is not $p$-trivial. Nevertheless, $(\C^*_p,d^*)$ has a
$p$-proximity order:

\begin{proposition}\label{lemma:quotient-space-ordering}
  For an extended quotient $(\C^*_p,d^*)$ of a Robinson space $(X,d)$
  one can compute a $p$-proximity order $\prec$ for some (unknown)
  compatible order $<$ of $(\C^*_p,d^*)$.
\end{proposition}

\begin{proof} By \Cref{extquotient}, $(\C^*_p,d^*)$ is Robinson. By
  \Cref{prop:proximity-preorder} there is a $p$-proximity pre-order
  $\preceq$ for a compatible order $<$ of $(\C^*_p,d^*)$. One can
  easily see that the copoints of $p$ in $(\C^*_p,d^*)$ are either
  trivial or of the form 
  $(C_i',C_i'')$ for a separable copoint $C_i \in \C_p$. We refine
  $\preceq$ into an order $\prec$ by arbitrarily ordering each such
  pair $\{C'_i, C''_i\}$ with $C'_i < C''_i$. We assert that there
  exists a compatible order $<^*$ on $\C^*_p$ having $\prec$ as a
  $p$-proximity order. By \Cref{extquotient}, $(\mathcal{C}^*_p,d^*)$
  is isometric to the restriction of $(X,d)$ to the following set: the
  point $p$, one representative $x_i$ for each tight or non-separable
  copoint $C_i$, and a diametral pair $\{x'_i,x''_i\}$ of $C_i$ as
  representatives for each separable copoint $C_i$ (we then have
  $x'_i < p < x''_i$ or $x''_i < p < x'_i$).

  We define $<^*$ from $<$ by permuting the representatives
  $(x'_i,x''_i)$ of each separable copoint $C_i$ with $x''_i < x'_i$,
  so that $x'_i <^* x''_i$ (in this case, $x'_i$ and $x''_i$ are said
  to be {\em permuted}). Then, $<^*$ is also a compatible order:
  indeed the dissimilarity matrices with rows and columns ordered by
  $<$ and $<^*$ are identical, because $\{x'_i,x''_i\}$ is an mmodule
  of $\C_p^*$. Next we prove that $\prec$ is a $p$-proximity order for
  $<^*$. To prove~\ref{item:proxorder1}, as $\prec$ is a refinement of
  $\preceq$, we only need to prove that $d(p, x'_i) \leq d(p, x''_i)$
  for any separable copoint. This is indeed so because
  $d(p, x'_i)=d(p, x''_i)$. To prove~\ref{item:proxorder2}, we first
  prove that $\preceq$ is a pre-order for $<^*$.

  We distinguish four cases. First, suppose that $x \precneq y$ and
  neither $x$ nor $y$ was permuted with each other or another point.
  Then the relative order of $p$, $x$ and $y$ does not change, hence
  in $<^*$, $y$ is not between $p$ and $x$. Second, suppose that
  $x \precneq y$, $x$ was permuted as member of a pair $(x'_i,x''_i)$,
  and $y$ was not permuted. This means that $x''_i < p < x'_i < y$
  (because $\preceq$ is a $p$-proximity pre-order), and after
  permuting we have $x'_i <^* p <^* x''_i <^* y$, hence $y$ is not
  between $x'_i$ and $p$ nor between $p$ and $x''_i$ in $<^*$. Now,
  suppose that $x \precneq y$ and $x$ was not permuted, but $y$ was
  permuted as member of a pair $(y'_j,y''_j)$. This is similar to the
  previous case: $y''_j < p < x < y'_j$ or $y''_j < x < p < y'_j$, and
  after permutation, we have $y'_j <^* p <^* x <^* y''_j$ or
  $y'_j <^* x <^* p <^* y''_j$. Anyways we get that $y$ is not between
  $p$ and $x$ in $<^*$. Finally, suppose that $x \precneq y$ and both
  $x,y$ were permuted. Hence they belong to pairs $(x'_i,x''_i)$ and
  $(y'_j, y''_j)$, respectively. Then
  $y''_j < x''_i < p < x'_i < y'_j$ and
  $y'_j <^* x'_i <^* p <^* x''_i <^* y''_j$. Hence $y$ is not between
  $p$ and $x$ in $<^*$. Consequently, we proved that $\preceq$ is a
  pre-order for $<^*$.

  Since $\prec$ is a refinement of $\preceq$, it suffices to
  prove~\ref{item:proxorder2} for the pairs $(x'_i,x''_i)$. But
  $x'_i <^* p <^* x''_i$, hence $x''_i$ is not between $p$ and $x'_i$,
  concluding the proof that $\prec$ is a $p$-proximity order for
  $<^*$.
\end{proof}

\subsection{Compatible orders from \texorpdfstring{$p$}{p}-proximity orders}\label{subs:co-from-proxor}

From the definition of a $p$-proximity order $\prec$ for a compatible
order $<$, it follows that $<$ can be recovered provided $\prec$ is
given and we know which elements of $X\setminus \{ p\}$ are located to
the left and to the right of $p$. This is specified by the following
result:

\begin{lemma}\label{remark:proxorder-bipartition-implies-compatorder}
  Let $(X,d)$ be a Robinson space with a compatible order $<$. Then $<$
  is fully determined by a $p$-proximity order $\prec$ for $<$ and the
  bipartition $(L,R) = (\{x \in X : x < p\}, \{x \in X : x > p\})$.
  More precisely, for any $u, v \in X \setminus \{p\}$, we have $u < v$ if and only
  if
  \begin{enumerate}[label=(\roman*),nosep]
  \item either $u \in L$, $v \in R$,
  \item or $u, v \in L$, and $v \prec u$,
  \item or $u, v \in R$ and $u \prec v$.
  \end{enumerate}
\end{lemma}

For a point $u \in X\setminus\{p\}$, we denote by $\side(u)$ the set
$L$ or $R$ to which $u$ belongs and by $\opposite(u)$ the other set.
Then we have the following elementary result:

\begin{lemma}\label{lemma:side-opposite}
  Let $(X,d)$ be a Robinson space, $p$ a point of $X$, $\prec$ a
  $p$-proximity order on $(X,d)$ and $u,v \in X\setminus\{p\}$ such
  that $u \prec v$. Then
  \begin{enumerate}[label=(\roman*),nosep]
  \item $d(u,v) < d(p, v)$ implies that $\side(u) = \side(v)$,
  \item $d(u,v) > d(p, v)$ implies that $\side(u) = \opposite(v)$.
  \end{enumerate}
\end{lemma}

It remains to construct the sets $L$ and $R$; this is done by
\Cref{algo:sortByBipartition}. To explain how this algorithm works we
will use two graphs $G$ and $H$; they are not explicitly used by
\Cref{algo:sortByBipartition} but are used in the proof of its
correctness. The graph $G$ has $X \setminus \{p\}$ as the vertex-set
and the edges are defined in such a way that the side of each vertex
in a connected component of $G$ is uniquely defined by fixing
arbitrarily the side of any vertex from that component. The second
graph $H$ has the connected components of $G$ plus $\{p\}$ as vertices
and the pairs of components which are ``tangled'' with respect to
$\prec$ as edges. Again, fixing the side of an arbitrary point from a
connected component of $H$, uniquely defines the side of all points of
$X\setminus \{ p\}$ belonging to that connected component. Finally, we
prove that if we order the connected components
${\mathcal{K}_0 = \{\{p\}\}, \mathcal K}_1,\ldots,{\mathcal K}_s$ of
$H$ by their maximal elements $m_1,\ldots, m_s$ in the order $\prec$,
then for any $i=1,\ldots, s-1$, the union
$\bigcup_{j=0}^i {\mathcal K}_i$ is an mmodule $(X,d)$ and an interval
of $\prec$. This implies that the sides of points in each
${\mathcal K}_i$ can be determined independently of the sides of the
points from ${\mathcal K}_{i+1},\ldots,{\mathcal K}_s$.

The graph $G$ has $X \setminus \{p\}$ as the vertex-set and
$\{ uv : u \prec v \land d(u,v) \neq d(p,v) \}$ as the edge-set.
Denote by $\mathcal{C}$ the set of connected components of the graph
$G$.

\begin{lemma}\label{connected-comp-G}
  For any connected component $C$ of the graph $G$, the sides of all
  points $C$ are uniquely determined by fixing arbitrarily the side of
  any vertex $r$ of $C$.
\end{lemma}

\begin{proof}
  Let $T$ be any spanning tree of $C$ and suppose that $T$ is rooted
  at the vertex $r$. Suppose also that $\side(r)$ was fixed. Then we
  proceed by induction on the length of the unique path of $T$
  connecting $r$ with any vertex $v\in C$. Let $uv$ be the edge of $T$
  on this path; whence, $\side(u)$ was already determined. Since $uv$
  is an edge of $G$, either we have $u\prec v$ and $d(u,v)\ne d(p,v)$
  or $v\prec v$ and $d(u,v)\ne d(p,u)$. In both cases, $\side(v)$ is
  well-defined by applying \Cref{lemma:side-opposite}.
\end{proof}

By \Cref{connected-comp-G}, if $G$ is connected, then the sides of all
vertices of $G$ are well-defined and one can retrieve the sets $L$ and
$R$ up to symmetry. Now suppose that $G$ is not connected. We say that
two connected components $C, C'\in \mathcal{C}$ of $G$ are
\emph{tangled} if $C$ and $C'$ are not comparable by $\prec$, i.e.,
either there exist $x,y\in C$ and $z\in C'$ such that
$x\prec z\prec y$ holds or there exist $x,y\in C'$ and $z\in C$ such
that $x\prec z\prec y$ holds.

\begin{lemma}\label{tangled}
  If $C,C' \in {\mathcal C}$ and $CC'$ is a tangled pair for which
  there exist $x,y \in C$ and $z \in C'$ such that $x \prec z\prec y$,
  then there exist $x',y'\in C$ such that $x' \prec z \prec y'$ and
  $x'y'$ is an edge of $G$.
\end{lemma}

\begin{proof}
  Since $x,y\in C$ there exists a path
  $P = (x=x_1,x_2,\ldots, x_{t-1},x_t=y)$ connecting the vertices $x$
  and $y$ in $C$. Since $x \prec z \prec y$, necessarily $P$ contains
  an edge $x_ix_{i+1}$ such that $x_i \prec z \prec x_{i+1}$. Thus we
  can set $x'=x_i$ and $y'=x_{i+1}$.
\end{proof}

\begin{lemma}\label{edge-tangled}
  Let $x \prec z \prec y$, $xy$ be an edge of $G$, and suppose
  that $yz$ is not an edge of $G$. If $d(x,y) < d(p,y)$, then
  $\side(x) = \side(y) \ne \side(z)$ and if $d(x,y) > d(p,y)$,
  then $\side(x) \ne \side(y) = \side(z)$.
\end{lemma}

\begin{proof}
  Since $x\prec y$ and $xy$ is an edge of $G$, we have
  $d(x,y)\ne d(p,y)$. On the other hand, since $zy$ is not an edge
  of $G$, we have $d(z,y) = d(p,y)$.

  First case: $d(x,y) < d(p,y)$. By \Cref{lemma:side-opposite}, we
  have $\side(x) = \side(y)$. Suppose that $\side(z) = \opposite(y)$
  in a compatible order $<$ for which $\prec$ is a $p$-proximity
  order. Since $x\prec z\prec y$, by
  \Cref{remark:proxorder-bipartition-implies-compatorder} $z$ must be
  located between $x$ and $y$ in $<$. Consequently,
  $d(z,y)=d(p,y)>d(x,y)$, contrary to the fact that in $<$ the point
  $z$ is located between $x$ and $y$. This contradiction shows that
  $\side(z)\ne \side(x)=\side(y)$ in this case.

  Second case: $d(x,y)>d(p,y)$. By \Cref{lemma:side-opposite}, we have
  $\side(x) \neq \side(y)$. Suppose that $\side(z)=\side(x)$. Since
  $x\prec z$, by
  \Cref{remark:proxorder-bipartition-implies-compatorder} $x$ must be
  located between $z$ and $p$ in any compatible order $<$ for which
  $\prec$ is a $p$-proximity order. Since
  $\side(z) = \side(x) = \opposite(y)$, $x$ is also located between
  $z$ and $y$ in $<$. Consequently, $d(x,y) > d(p,y) = d(z,y)$, which
  is impossible because in $<$ the point $x$ is located between $z$
  and $y$. This contradiction shows that $\side(z)=\side(y)$ in this
  case. 
\end{proof}

\begin{lemma}\label{tangled-side}
  If $C,C'\in {\mathcal C}$ and $CC'$ is a tangled pair, then the
  sides of all points of $C'$ are uniquely determined by the sides of
  points of $C$ and, vice-versa, the sides of all points of $C$ are
  uniquely determined by the sides of points of $C'$.
\end{lemma}

\begin{proof}
  Since $CC'$ is a tangled pair, either there exist $x,y\in C$ and
  $z\in C'$ such that $x\prec z\prec y$ or there exist $x,y\in C'$ and
  $z\in C$ such that $x\prec z\prec y$, say the first. By
  \Cref{tangled}, there exists an edge $x'y'$ of $C$ such that
  $x'\prec z\prec y'$. By \Cref{edge-tangled}, the side of $z$ is
  uniquely determined by the sides of $x'$ and $y'$ (and thus the
  sides of $x'$ and $y'$ are uniquely determined by the side of $z$).
  By \Cref{connected-comp-G} the sides of points of $C'$ are
  uniquely determined by the side of $z$.
\end{proof}

The graph $H$ has the set $\mathcal{C}$ of connected components of $G$
as vertices and the tangled pairs of $\mathcal{C}$ as edges. Let
$\Sigma=\{ {\mathcal K}_1,\ldots, {\mathcal K}_s\}$ be the connected
components of the graph $H$. Set also ${\mathcal K}_0:=\{ p\}$. We
will denote by ${\mathcal K}^{\oplus}_i$ the set of all points of $X$
belonging to the connected components of $G$ included in
${\mathcal K}_i$ and by ${\mathcal K}^{\oplus}_{\le i}$ the union of
the sets ${\mathcal K}^{\oplus}_0,$ ${\mathcal K}^{\oplus}_1,\ldots,$
${\mathcal K}^{\oplus}_i$. For each ${\mathcal K}_i, i=1,\ldots,s$ we
denote by $m_i$ the maximal point of ${\mathcal K}^{\oplus}_i$ in
$\prec$. Let also $C(m_i) \in \mathcal{C}$ be the component of $G$
containing the point $m_i$. Suppose that the connected components of
$H$ are ordered $\{ {\mathcal K}_1,\ldots, {\mathcal K}_s\}$ according
to the order of the points $m_1,\ldots, m_s$ in $\prec$:
$p\prec m_1\prec\ldots\prec m_s$.

\begin{lemma}\label{lemma:graphHisNotConnected}
  For any $i=1,\ldots,s$,
  ${\mathcal K}^{\oplus}_i=\{ x\in X\setminus \{ p\}: m_{i-1}\prec x\preceq m_i\}$
  holds and the interval
  $\{ x\in X: x\preceq m_i\}={\mathcal K}^{\oplus}_{\le i}$ is an
  mmodule of $(X,d)$.
\end{lemma}

\begin{proof}
  It suffices to establish the lemma for the last component
  ${\mathcal K}_s$ and use induction. Let $u_*$ be the minimal element
  of $\mathcal{K}^{\oplus}_s$ for $\prec$. By definition, $m_{s-1}$ is
  the maximal element of ${\mathcal K}^{\oplus}_{\le s-1}$. Therefore
  to prove that
  ${\mathcal K}^{\oplus}_s=\{ x\in X\setminus \{ p\}: m_{s-1}\prec x\preceq m_s\}$
  it suffices to prove that $m_{s-1}\prec u_*$. Suppose, by a way of
  contradiction, that $m_{s-1} > u_*$. Let $C'$ be the connected
  component of $G$ containing $u_*$. Since $C(m_s)$ and $C'$ belong to
  $\mathcal{K}_s$, there exists a path
  $C'= C_0, C_1,\ldots, C_{k-1}, C_k=C(m_s)$ connecting $C'$ and
  $C(m_s)$ in the graph $H$. Let $w_i, t_i$ be respectively the
  minimal and the maximal elements of $C_i$ with respect to $\prec$.
  Since for all $0\leq i<k$, $C_iC_{i+1}$ are tangled pairs, in
  $\prec$ the intervals $[w_i,t_i]$ and $[w_{i+1},t_{i+1}]$ have a
  nonempty intersection. Since $w_0\prec m_{s-1}\prec t_s$,
  necessarily there exists an index $i$ such that
  $w_i \prec m_{s-1}\prec t_i$. This implies that the connected
  components $C(m_{s-1})$ and $C_i$ are tangled. By the definition of
  the graph $H$, $C(m_{s-1})$ must belong to $\mathcal{K}_s$, a
  contradiction with the definition of the point $m_{s-1}$. This shows
  that $m_{s-1}\prec u_*$ and establishes the first assertion for
  ${\mathcal K}_s$. Applying induction we conclude that
  ${\mathcal K}^{\oplus}_i=\{ x\in X\setminus \{ p\}: m_{i-1}\prec x\preceq m_i\}$
  holds for any $i=1,\ldots,s$. This also implies that
  ${\mathcal K}^{\oplus}_{\le i}=\{ x\in X: x\preceq m_i\}$ holds for
  all $i=1,\ldots,s$.

  To prove that ${\mathcal K}^{\oplus}_{\le i}$ is an mmodule, again
  we will prove the assertion for $i=s-1$ and use induction to get the
  result for all $i$. Let $M=\{ x\in X: x\preceq m_{s-1}\}$. Let
  $x,y \in M$ and $q \notin M$. Then $x \prec q$ and $y \prec q$.
  Since $x$ and $y$ do not belong to the connected component of $G$
  containing $q$, we also have $d(p,q) = d(q,x)$ and
  $d(p,q) = d(q,y)$. Hence $d(q,x) = d(q,y)$, proving that $M$ is an
  mmodule.
\end{proof}

From \Cref{tangled} it follows that a bipartition $(L_i,R_i)$ of
points of each connected component ${\mathcal K}_i$ of the graph $H$
is uniquely determined once the side of one of its points, say of
$m_i$, is arbitrarily fixed. Let $L = \bigcup_{i=1}^s L_i$ and
$R = \bigcup_{i=1}^s R_i$. Now, we prove that the total order obtained
from the bipartition $(L,R)$, where $L$ and $R$ are ordered according
to the $p$-proximity order $\prec$ (see
\Cref{remark:proxorder-bipartition-implies-compatorder}) is a
compatible order.

\begin{proposition}\label{compatiblefromproximity}
  Let $(X,d)$ be a Robinson space and $\prec$ be a $p$-proximity
  order. Let $<'$ be a total order obtained from the bipartition
  $(L,R)$ according to $\prec$, where
  $L=\bigcup_{i=1}^s L_i, R=\bigcup_{i=1}^s R_i$ and $(L_i,R_i)$ is
  the bipartition of ${\mathcal K}^{\oplus}_{i}$ obtained by
  arbitrarily fixing $\side(m_i)\in \{ L,R\}$. Then $<'$ is a
  compatible order of $(X,d)$.

  Conversely, any compatible order $<$ of $(X,d)$ such that $\prec$ is
  a $p$-proximity order for $<$ is obtained in this way. Consequently,
  $\prec$ is a $p$-proximity order for $2^s$ compatible orders.
\end{proposition}

\begin{proof}
  Let $<$ be a compatible order for which $\prec$ is a $p$-proximity
  order. We prove that $<'$ is a compatible order by induction on $s$.
  If $s=1$, then the graph $H$ is connected and there are only two
  partitions $(L,R)$ and $(R,L)$ and $<'$ coincides with $<$ or with
  its opposite, thus $<'$ is compatible on
  $\{ p\}\cup {\mathcal K}^{\oplus}_s$. For the same reason, for any
  connected component ${\mathcal K}_i$ of the graph $H$, the
  restriction of $<'$ on $\{ p\}\cup {\mathcal K}^{\oplus}_i$ is also
  compatible. Suppose by the induction hypothesis that the restriction
  of $<'$ on the union ${\mathcal K}^{\oplus}_{\le s-1}$ of the first
  $s-1$ connected components is a compatible order. To prove that $<'$
  is compatible on $X$, pick any three points $u,v,w$ such that
  $u <' v <' w$. We assert that $d(u,w) \geq \max \{d(u,v), d(v,w)\}$.
  If $\{ u,v,w\} \subseteq {\mathcal K}^{\oplus}_{\le s-1}$ or
  $\{ u,v,w\} \subseteq {\mathcal K}^{\oplus}_{s}$ the result follows
  by induction hypothesis or from the basic case.

  First, let $u,v\in {\mathcal K}^{\oplus}_{\le s-1}$ and
  $w\in {\mathcal K}^{\oplus}_{s}$. By
  \Cref{lemma:graphHisNotConnected}, ${\mathcal K}^{\oplus}_{\le s-1}$
  is an mmodule, thus $d(u,w)=d(v,w)$. Since $u \prec w$ and
  $v \prec w$, $w$ is not between $p$ and $u$ nor between $p$ and $v$
  in $<$, hence $w$ is not between $u$ and $v$ in $<$. Since $<$ is a
  compatible order, independently of the position of $w$, we get
  $d(u,v) \leq \max \{ d(u,w), d(v,w)\} = d(u,w)$ and we are done. The
  analysis of case $v, w \in {\mathcal K}^{\oplus}_{\le s-1}$ and
  $u \in {\mathcal K}^{\oplus}_{s}$ is analogous.

  The case $u, w \in {\mathcal K}^{\oplus}_{\le s-1}$ and
  $v \in {\mathcal K}^{\oplus}_{s}$ cannot happen, as
  $\mathcal{K}^\oplus_{\leq s-1}$ is an interval of $<'$.


  Next, let $v \in {\mathcal K}^{\oplus}_{\le s-1}$ and
  $u, w \in {\mathcal K}^{\oplus}_{s}$. Then because
  $\mathcal{K}^\oplus_{\leq s-1}$ is an interval of $<'$ containing
  $p$, $\side(u) = L$ and $\side(w) = R$. Since
  ${\mathcal K}^{\oplus}_{\le s-1}$ is an mmodule containing $p$ and
  $v$, we deduce that $d(p,u) = d(v,u)$ and $d(p,w) = d(v,w)$. Since $<'$
  is compatible on $\{p\} \cup {\mathcal K}^{\oplus}_{s}$ and
  $u <' p <' w$, we get $d(u,w) \geq \max\{ d(u,p), d(p,w)\}$. Combining
  these equalities and inequalities we obtain that
  $d(u,w) \ge \max\{ d(u,v),d(v,w)\}$.

  Finally, suppose that $w \in {\mathcal K}^{\oplus}_{\le s-1}$ and
  $u, v \in {\mathcal K}^{\oplus}_{s}$. Since
  $\mathcal{K}^\oplus_{\leq s-1}$ is an interval of $<'$ containing
  $p$, $\side(u) = \side(v) = L$. Since
  ${\mathcal K}^{\oplus}_{\le s-1}$ is an mmodule containing $p$ and
  $w$, we have $d(p,u) = d(w,u)$ and $d(p,v) = d(w,v)$. Since $<'$ is
  compatible on $\{p\} \cup {\mathcal K}^{\oplus}_{s}$ and
  $u <' v <' p$, we have $d(u,p) \geq \max \{ d(u,v),d(v,p)\}$.
  Combining these equalities and inequalities we obtain that
  $d(u,w) \geq \max \{d(u,v), d(v,w)\}$. The case when
  $u \in \mathcal{K}^\oplus_{\leq s-1}$,
  $v,w \in \mathcal{K}^\oplus_s$ is symmetrical. This concludes the
  proof of the first assertion.

  To prove the second assertion it suffices to show how the compatible
  order $<$, for which $\prec$ is a $p$-proximity order, can be
  obtained by our construction. Clearly, for each connected component
  ${\mathcal K}_i$ of the graph $H$, the point $m_i$ is located either
  to the left or to the right of $p$. Therefore $\side(m_i)$ can be
  fixed in this way and the bipartition $(L_i,R_i)$ of
  ${\mathcal K}^{\oplus}_{i}$ is uniquely determined by the value of
  $\side(m_i)$. Moreover, this bipartition coincides with the
  bipartition of ${\mathcal K}^{\oplus}_{i}$ with respect to the order
  $<$. By \Cref{remark:proxorder-bipartition-implies-compatorder} $<$
  coincides with the total order defined by the bipartition $(L,R)$,
  where $L = \bigcup_{i=1}^s L_i, R = \bigcup_{i=1}^s R_i$.
\end{proof}

Summarizing the previous results, we obtain the following simple
algorithm for computing a compatible order of $(X,d)$ from a
$p$-proximity order $\prec$:

\begin{enumerate}[label=\arabic*.]
\item Compute the graph $G$ and the set $\mathcal C$ of its connected
  components,
\item Compute the graph $H$ and the set
  $\Sigma=\{ {\mathcal K}_1,\ldots, {\mathcal K}_s\}$ of its connected
  components,
\item Compute the rightmost point $m_i$ of each
  ${\mathcal K}^{\oplus}_i$, $i=1,\ldots,s$ with respect to $\prec$
  and arbitrarily fix $\side(m_i)\in \{ L,R\}$, $i=1,\ldots,s$
\item Using $\side(m_i)$, derive $\side(x)$ for all other
  points $x\in {\mathcal K}^{\oplus}_i$ for $i=1,\ldots,s$,
\item Return the total order $<'$ on $(X,d)$ in which the points of
  $\{ x\in X\setminus \{ p\}: \side(x)=L\}$ and of
  $\{ x\in X\setminus \{ p\}: \side(x)=R\}$ are ordered according to
  $\prec$.
\end{enumerate}

The complexity of this algorithm is $O(|X|^2)$ since the complexity of
each of its steps is $O(|X|^2)$. The unique step requiring some
explanation is the computation of $H$. For this, for each connected
component $C_i$ of $G$ we compute the minimum $d_i$ and the maximum
$e_i$ according to $\prec$. Then we sort these segments $[d_i,e_i]$ by
the ending dates and sweep the sorted list to return the pairs of
intersecting segments $[d_i,e_i]$ and $[d_j,e_j]$. This corresponds to
the edges of $H$ and all this can be done in $O(|X|^2)$.

In fact, to compute the sides of points and the resulting compatible
order, we do not need to explicitly compute the graphs $G$ and $H$.
This leads us to \Cref{algo:sortByBipartition}. Actually, this
algorithm simultaneously constructs the connected components
$\mathcal{K}_i$ of $H$ (without building all the components of the
graph $G$) and assigns the points of each $\mathcal{K}^{\oplus}_i$ to
the correct set $L$ or $R$.

\begin{algorithm}[th]
  \caption{$\textrm{sortByBipartition}(p,X)$}
  \label{algo:sortByBipartition}
  \begin{algorithmic}[1]
    \Require{A Robinson space $(X,d)$ ($d$ is implicit), a point $p \in X$, $X\setminus \{p\}$ is given in $p$-proximity order.}
    \Ensure{$X$ in a compatible order.}
      \Let $L = \emptylist$, $R = \emptylist$, $\variable{Undecided} = \textrm{reverse}(X\setminus\{p\})$
      \For{$q \in X\setminus \{p\}$ in decreasing order} \label{step:sortByBipartition:beginMainLoop}
        \If{$q \in \variable{Undecided}$} \label{step:sortBiPartart:testGeneral}
          \State choose arbitrarily: either $R \gets q \cons R$ or $L \gets q \cons L$ \label{step:sort-bi-part:choice-arbitrar}
          \State $\variable{Undecided} \gets \variable{Undecided} \setminus \{q\}$
        \EndIf
        \Let $\variable{Skipped} = []$
        \For{$x \in \variable{Undecided}$ from first to last} \label{step:sortByBipart:begin-loop}
          \If{$d(x,q) = d(p,q)$}
            \State $\variable{Skipped} \gets x \cons \variable{Skipped}$ \label{step:sort-bi-part:equal-skipped}

          \Else \label{step:sortBiPart:x-nonEqu-q}
          \If {($d(x,q) < d(p,q)$ and $q \in L$) or ($d(x,q) > d(p,q)$ and $q \in R$)} \label{step:sortBiPart:x-near-q}
            \State $L \gets x \cons L$ \label{step:sortBiPart:sameSide}
            \State $R \gets \variable{Skipped} \append R$ \label{step:sortBiPart:sameSide-Bis}
          \Else
            \State $R \gets x \cons R$ \label{step:sortBiPart:oppositeSide}
            \State $L \gets \variable{Skipped} \append L$  \label{step:sortBiPart:oppositeSide-Bis}
          \EndIf
          \State $\variable{Skipped} \gets []$  \label{step:sortByBipart:end-loop}
          \EndIf
        \EndFor
        \State $\variable{Undecided} \gets \textrm{reverse}(\variable{Skipped})$ \label{step:sort-bi-part:undecided}
      \EndFor
      \Return $\textrm{reverse}(L) \append [p] \append R$
    \end{algorithmic}
\end{algorithm}

\begin{proposition}\label{prop:algo-sortByBipartition-correct}
  Given a Robinson space $(X,d)$ with $|X|=n$, a point $p\in X$, and a
  $p$-proximity order on $X\setminus\{p\}$,
  \Cref{algo:sortByBipartition} returns $X$ sorted along a compatible
  order in $O(n^2)$ time.
\end{proposition}

\begin{proof}
  We first prove the correctness of \Cref{algo:sortByBipartition}. We
  start by stating some invariants of the algorithm that holds at the
  end of each iteration of any loop, and can be readily checked. We
  consider that the loop at Line~\ref{step:sortByBipart:begin-loop}
  removes the element $x$ from $\variable{Undecided}$ at each
  iteration.

  \begin{enumerate}[label=(\roman*),ref=(\roman*)]
  \item\label{inv1} The list $\variable{Undecided}$ is always sorted
    in decreasing $p$-proximity order,
  \item\label{inv2} $\variable{Skipped}$, $\variable{L}$ and
    $\variable{R}$ are sorted in increasing $p$-proximity order.
  \item\label{inv3} each element is in exactly one of the four lists
    $\variable{Undecided}$, $\variable{Skipped}$, $\variable{L}$ and
    $\variable{R}$.
  \item\label{inv4} for any $x \in \variable{Undecided}$,
    $y \in \variable{Skipped}$ and
    $z \in \variable{L} \cup \variable{R}$, we have
    $x \prec y \prec z$.
  \end{enumerate}

  Obviously $\variable{L}$ and $\variable{R}$ contain the elements of
  each side, while $\variable{Undecided}$ and $\variable{Skipped}$
  contains elements whose side is not determined yet.

  Notice that the first element $q$ for which
  Line~\ref{step:sort-bi-part:choice-arbitrar} is executed is $m_s$.
  Consider the iterations of
  Loop~\ref{step:sortByBipartition:beginMainLoop}--\ref{step:sort-bi-part:undecided}
  starting from the first one and as long as
  Line~\ref{step:sort-bi-part:choice-arbitrar} is not executed again. We
  claim that at the end of these iterations, $\variable{Undecided}$
  contains exactly $\mathcal{K}^\oplus_{\leq s-1}$, while
  $\variable{L} \cup \variable{R}$ contains $\mathcal{K}^\oplus_s$. We
  denote $S \subseteq X \setminus \{p\}$ the set of elements in
  $\variable{L} \cup \variable{R}$ at that time. First we prove that
  if a component $C$ of $G$ intersects $S$, then $C \subseteq S$, in
  particular $C(m) \subseteq S$. Suppose not, then there is an edge
  $uv$ in $G$ with $u,v \in C$, $u \notin S$ and $v \in S$, and by
  invariant~\ref{inv4} $u \prec v$. But then during the
  Loop~\ref{step:sortByBipartition:beginMainLoop}--\ref{step:sort-bi-part:undecided}
  for $q = v$ and
  Loop~\ref{step:sortByBipart:begin-loop}--\ref{step:sortByBipart:end-loop}
  for $x = u$, $u$ should have been decided on
  Lines~\ref{step:sortBiPart:x-near-q}--\ref{step:sortByBipart:end-loop},
  contradiction. Then we prove that if $CC'$ is a tangled pair with
  $u,v \in C$, $w \in C'$, $u \prec w \prec v$, $C \subseteq S$, then
  $C' \subseteq S$. This follows immediately from
  Invariant~\ref{inv4}. Finally, as $G$ has no edge between
  $\mathcal{K}^\oplus_{\leq s-1}$ and $\mathcal{K}^\oplus_s$, when
  $x$ is an element of $\mathcal{K}^\oplus_{\leq s-1}$,
  $d(x,q) = d(p,q)$ during these iterations, hence $x$ is skipped
  (Line~\ref{step:sort-bi-part:equal-skipped}), proving our claim.

  From this we proceed by induction on $s$. It only remains to prove
  that each element is correctly assigned to its side. On
  Line~\ref{step:sort-bi-part:choice-arbitrar}, it follows from the
  fact that $q = m_i$, hence we can choose arbitrarily by
  \Cref{compatiblefromproximity}. On
  Lines~\ref{step:sortBiPart:sameSide}
  and~\ref{step:sortBiPart:oppositeSide}, it follows from
  \Cref{lemma:side-opposite}. On
  Lines~\ref{step:sortBiPart:sameSide-Bis}
  and~\ref{step:sortBiPart:oppositeSide-Bis}, it follows from
  \Cref{edge-tangled}. This proves the correction of
  \Cref{algo:sortByBipartition}.

  The complexity of \Cref{algo:sortByBipartition} is easily derived,
  since each loop iterates at most $n$ times and by observing that the
  complexity of appending $\variable{Skipped}$ is amortized over the
  insertions into $\variable{Skipped}$.
\end{proof}

Consequently, we obtain the following result:

\begin{proposition}\label{proximity->compatible}
  Let $(X,d)$ be $p$-trivial or flat Robinson space on $n$ points.
  Then a compatible order on $X$ can be computed in $O(n^2 + T)$ time,
  where $T$ is the total time used by the $\variable{refine}$
  procedure. Analogously, if $(\C^*_p,d^*)$ is an extended quotient of
  a Robinson space $(X,d)$ with $k$ copoints, then a compatible order
  on $\C^*_p$ can be computed in $O(k^2 + T)$.
\end{proposition}

\begin{proof}
  If $(X,d)$ is $p$-trivial, then the result follows from
  \Cref{prop:proximity-order,prop:algo-sortByBipartition-correct}.
  Now suppose that $(X,d)$ is flat and let $p$ be a diametral point of
  $(X,d)$. By \Cref{prop:mmodules-in-unique-order}, $(X,d)$
  is $p$-trivial, thus we can apply the previous case. Finally, if
  $(\C^*_p,d^*)$ is an extended quotient of a Robinson space $(X,d)$,
  then $C^*_p$ contains at most $2k$ points. Consequently the result
  follows from
  \Cref{lemma:quotient-space-ordering,prop:algo-sortByBipartition-correct,remark:proxorder-bipartition-implies-compatorder}.
\end{proof}

\begin{runningexample}
  We illustrate \Cref{algo:sortByBipartition} on some subspace of the
  running example. Consider the $1$-proximity order
  $$1 \prec 17 \prec 9 \prec 10 \prec 6 \prec 4 \prec 13 \prec 7 \prec 19$$
  for which we want to find a compatible order, with dissimilarities
  given in \Cref{TABLE:example-sbb} (left).

  \begin{figure}
    {\footnotesize
    \begin{displaymath}
      \begin{array}{cp{0.1cm}ccccccccc}
           & &  1 & 17 &  9 & 10 &  6 &  4 & 13 &  7 & 19 \vspace{0.1cm}\\
         1 & &  0 &  4 &  4 &  8 &  8 &  9 &  9 &  9 & 10 \\
        17 & &    &  0 &  6 &  7 &  8 &  9 &  9 &  9 & 10 \\
         9 & &    &    &  0 &  8 &  9 &  9 &  9 &  9 & 10 \\
        10 & &    &    &    &  0 &  7 &  9 &  9 &  9 & 10 \\
         6 & &    &    &    &    &  0 &  9 &  9 &  9 & 10 \\
         4 & &    &    &    &    &    &  0 &  6 &  9 &  8 \\
        13 & &    &    &    &    &    &    &  0 &  9 &  5 \\
         7 & &    &    &    &    &    &    &    &  0 & 11 \\
        19 & &    &    &    &    &    &    &    &    &  0 \\
      \end{array}
      \qquad\qquad
      \begin{array}{cp{0.1cm}ccccccccc}
           & & 19 & 13 &  4 &  9 &  1 & 17 & 10 &  6 &  7 \vspace{0.1cm}\\
        19 & &  0 &  5 &  8 & 10 & 10 & 10 & 10 & 10 & 11 \\
        13 & &    &  0 &  6 &  9 &  9 &  9 &  9 &  9 &  9 \\
         4 & &    &    &  0 &  9 &  9 &  9 &  9 &  9 &  9 \\
         9 & &    &    &    &  0 &  4 &  6 &  8 &  9 &  9 \\
         1 & &    &    &    &    &  0 &  4 &  8 &  8 &  9 \\
        17 & &    &    &    &    &    &  0 &  7 &  8 &  9 \\
        10 & &    &    &    &    &    &    &  0 &  7 &  9 \\
         6 & &    &    &    &    &    &    &    &  0 &  9 \\
         7 & &    &    &    &    &    &    &    &    &  0 \\
      \end{array}
    \end{displaymath}
    }
    \caption{An extended quotient space from the running example,
      given in $1$-proximity order (left) and compatible order
      (right).}
    \label{TABLE:example-sbb}
  \end{figure}

  First, we consider $q = 19$, with $d(p,q) = 10$, on
  Line~\ref{step:sort-bi-part:choice-arbitrar}, say we choose
  $19 \in L$. Then for $x = 7$, $d(q,x) = 11 > d(p,q) = 10$, hence
  $7 \in R$. Similarly, for $x = 13$ and $x=4$, the algorithm decides
  that $13 \in L$, $4 \in L$. Then all remaining elements are skipped.

  Then for $q \in \{7, 13, 4\}$ all remaining elements are again
  skipped. When $q = 6$, we get to choose arbitrarily on
  Line~\ref{step:sort-bi-part:choice-arbitrar} that $6 \in R$. Then
  $d(p,q) = 8$. For $x = 10$, $d(q,x) = 7 < d(p,q)$ implies
  $10 \in R$, for $x = 9$, $d(q,x) = 9 > d(p,q)$ implies $9 \in L$,
  and for $x = 17$, $d(q,x) = 8 = d(p,q)$, $17$ is skipped.

  Then for $q = 10$, $d(p,q) = 8 > d(17,q)$, hence $17 \in R$. The
  algorithm returns the compatible order
  $$19 < 13 < 4 < 9 < 1 < 17 < 10 < 6 < 7,$$
  whose matrix is given in \Cref{TABLE:example-sbb} (right).
\end{runningexample}

\section{A divide-and-conquer algorithm}\label{s:divide-and-conquer-copoint}

In this section, we describe the divide-and-conquer algorithm for
recognizing Robinson spaces, prove its correctness and establish its
running time.

\subsection{The algorithm}
The results of \Cref{s:mmodules,s:mmodulesRobinson,s:flat} (namely,
\Cref{location-copoints}, \Cref{quotient+copoints},
\Cref{proximity->compatible} and \Cref{algo:recursiveRefine},
\Cref{algo:separateIfSeparable}, \Cref{algo:sortByBipartition}) lead
to the following algorithm for computing a compatible order of a
Robinson space $(X,d)$:

\begin{enumerate}[label=\arabic*.]
\item Compute a copoint partition $\C_p$ of $(X,d)$ using
  \Cref{algo:recursiveRefine},
\item Recursively find a compatible order $<_i$ for each copoint $C_i$
  of $\C_p$,
\item Classify the copoints of $\C_p$ into separable, tight, and
  non-separable, separate the separable copoints using
  \Cref{algo:separateIfSeparable}, and construct the extended quotient
  $(\C^*_p,d^*)$ of $(X,d)$,
\item Compute a $p$-proximity order $\prec$ for the extended quotient
  $(\C^*_p,d^*)$ using \Cref{algo:sortByBipartition},
\item Build a compatible order $<^*$ for $(\C^*_p,d^*)$ using $\prec$,
\item Merge the compatible order $<^*$ on $\C^*_p$ with the compatible
  orders $<_i$ on the copoints $C_i$ of $\C_p$ to get a total order
  $<$ on $X$, using \Cref{quotient+copoints},
\item If $<$ is not a compatible order of $(X,d)$, then return ``not
  Robinson'', otherwise return $<$.
\end{enumerate}
The pseudo-code of the algorithm is \Cref{algo:findCompatibleOrder}.
\begin{algorithm}[!th]
  \caption{$\textrm{findCompatibleOrder(X)}$}
  \label{algo:findCompatibleOrder}
  \begin{algorithmic}[1]
  \Require{a Robinson space $(X,d)$ ($d$ is implicit).}
  \Ensure{a compatible order for $X$ (as a sorted list).}
    \If {$X$ is empty}
      \Return $[]$
    \EndIf
    \Let $p \in X$, $X' = X \setminus \{p\}$ \label{step:choice-p}
    \Let $[C_1,\ldots,C_k] = \textrm{recursiveRefine}(p, [p], X', [])$ \label{step:call-rec-Refine}
    \Let $\variable{representedCopoints} = []$
    \For {$i \in \{1,\ldots,k\}$ in decreasing order} \label{step:begin-loop}
      \Let $C'_i = \textrm{findCompatibleOrder}(C_i)$
      \State $\variable{representedCopoints} \gets \textrm{separateIfSeparable}(p,C'_i) \append \variable{representedCopoints}$ \label{step:end-loop}
    \EndFor
    \Let $[(x_1,T_1),\ldots,(x_{k'},T_{k'})] = \variable{representedCopoints}$
    \Let $[x_{\sigma(1)},\ldots,p,\ldots,x_{\sigma(k')}] = \textrm{sortByBipartition}(p, [x_1,\ldots,x_{k'}])$ \label{step:call-bi-partition}
    \Return $\textrm{concatenate}(T_{\sigma(1)}, \ldots, [p], \ldots, T_{\sigma(k')})$ \label{step:return}
  \end{algorithmic}
\end{algorithm}
%


To represent the extended quotient $(\C^*_p,d^*)$, we select a set of
representatives of tight, non-separable, and halved copoints: the
point $p$, one representative $x_i$ for each tight or non-separable
copoint $C_i$, and a diametral pair $(x'_i,x''_i)$ for each separable
copoint.

\subsection{Complexity and correctness of the algorithm} The
correctness and the complexity of \Cref{algo:sortByBipartition}
($\variable{sortByBipartition}$), \Cref{algo:separateIfSeparable}
($\variable{separateIfSeparable}$) and \Cref{algo:recursiveRefine}
($\variable{recursiveRefine}$) was established in
\Cref{subs:co-from-proxor,subs:conical,section:pProximityOrder}. For
\Cref{algo:recursiveRefine}, the complexity was established without
counting the calls of \Cref{algo:refine} ($\variable{refine}$). This
will be done here.

We will use the following auxiliary result:

\begin{lemma}\label{lemma:refinement-complexity}
  If $T : \mathbb{N} \to \mathbb{N}$ satisfies the recurrence relation
  $T(n) \leq \sum_{i=1}^k T(n_i) + n \log k$, for all partitions
  $\sum_{i=1}^k n_i = n$ of $n$ in $k \geq 2$ positive integers, then
  $T(n) = O(n^2)$.
\end{lemma}

\begin{proof}
  By convexity of the function $\sum_{i=1}^k x^2_i$, the maximum of
  $\sum_{i=1}^k n_i^2$ over all partitions of $n$ in $k$ parts is
  attained by a partition with one class with $n-k+1$ points and $k-1$
  singletons, and has value $(n-k+1)^2 + (k-1)$. Assume that for all
  $p < n$, $T(p) \leq \alpha p^2$ for some $\alpha \geq 1$. Then
  \begin{align*}
    T(n) &\leq \sum_{i=1}^k \alpha n_i^2 + n \log k \\
         &\leq \alpha (n-k+1)^2 + \alpha (k-1) + n \log k\\
         &= \alpha n^2 - \alpha (k - 1) (2n - k) + n \log k \\
         &\leq \alpha n^2 - n ( \alpha (k-1) - \log k)
  \end{align*}
  where the last inequality follows from $2n - k \geq n$. It suffices
  to prove that $\alpha (k-1) - \log k$ is nonnegative, which is true
  because $\alpha \geq 1$ and $k \geq 2$.
\end{proof}

We continue with the main result of the paper.

\begin{theorem}
  \Cref{algo:findCompatibleOrder} computes a compatible order
  of a Robinson space $(X,d)$ in $O(n^2)$ time.
\end{theorem}

\begin{proof}
  The correction follows from \Cref{quotient+copoints} that
  proves that a compatible order on $(X,d)$  can be built by composing a
  compatible order on each copoint or halved copoint with a compatible
  order on the extended quotient space (which exists by \Cref{quotient+copoints}).
  By \Cref{prop:separateIfSeparableComplexity} and by induction,
  each $T_i$ is a tight or non-separable copoint or a halved copoint in
  increasing compatible order, with representative $x_i$, and by
  \Cref{prop:algo-sortByBipartition-correct}, $\sigma$ sorts the
  representatives $(x_i)_{i \in \{1,\ldots,k\}}$ and $p$ in a compatible
  order, so that \Cref{quotient+copoints} applies.

  We analyze the complexity of \Cref{algo:findCompatibleOrder} by
  counting separately the number of operations done in the procedures
  $\mathit{refine}$, $\mathit{recursiveRefine}$ and
  $\mathit{sortByBipartition}$. All the other operations can be done
  in linear-time at each level of recursion, thus in $O(|X|^2)$ times
  in total.

  \begin{itemize}
  \item $\mathit{recursiveRefine}$ contributes $O(|X|^2)$ in the total
    complexity; indeed, applying
    \Cref{lemma:recursiveRefine-complexity}, the first call takes
    $\alpha \left(|X|^2 - \sum_{i=1}^k |C_i|^2\right)$ (for some constant
    $\alpha$), while the cost of $\mathit{recursiveRefine}$ in the
    recursive calls are at most $\alpha \sum_{i=1}^k |C_i|^2$ by
    induction, summing to $\alpha |X|^2$.

  \item $\mathit{refine}$ contributes $O(|X|^2)$ in the total
    complexity, because it follows the recurrence relation described in
    \Cref{lemma:refinement-complexity}.

  \item $\mathit{sortByBipartition}$ contributes $O(|X|^2)$ in the
    total complexity; indeed, considering the recursion tree of calls
    to $\mathit{findCompatibleOrder}$, one can see that each call to
    $\mathit{sortByBipartition}$ uses $O(k^2)$ operations where $k$ is
    the arity of the corresponding node. Hence the contribution of
    $\mathit{sortByBipartition}$ is of the form
    $\beta \sum_{i = 1}^l k_i^2$ where $l$ is the number of nodes and
    $k_i$ the arity of the $k_i$th node. But, because each node, inner
    or leaf, can be associated to a unique element in $X$,
    $\sum_{i=1}^l k_i = |X| - 1$, implying that
    $\beta \sum_{i=1}^l k_i^2 \leq \beta |X|^2$ by convexity.
  \end{itemize}

  Summing up all the contributions, we get that
  $\variable{findCompatibleOrder}$ runs in time $O(|X|^2)$.
\end{proof}

\begin{remark} \Cref{algo:findCompatibleOrder} can be transformed into
  a recognition algorithm by simply testing in $O(|X|^2)$ time if the
  returned sorted list is a compatible order on $(X,d)$. If this is
  not the case, from the results of previous sections it follows that
  $(X,d)$ is not a Robinson space.
\end{remark}

\begin{remark}
  If $(X,d)$ is $p$-trivial, then all copoints $C_i$ have size 1, thus
  in this case the \Cref{algo:findCompatibleOrder} is no longer
  recursively applied to the copoints. In particular, this is the case
  if $(X,d)$ is a flat Robinson space and $p$ is diametral, since by
  \Cref{prop:mmodules-in-unique-order} $(X,d)$ is then $p$-trivial.
\end{remark}

\begin{runningexample}
  We conclude this section by running $\variable{findCompatibleOrder}$
  on the running example. On Line~\ref{step:choice-p}, we chose
  $p = 1$ and then on Line~\ref{step:call-rec-Refine} we build the
  ordered copoint partition
  $$\mathcal{C}_{1}=[\{1\}, \{17\}, \{9\}, \{10\}, \{6\}, \{3,4,8,16,18\},\{11,13,14\}, \{7\}, \{2,5,12,15,19\} ]$$
  by using $\mathit{recursiveRefine}$ (see
  \Cref{section:pProximityOrder}). It returns the following orders of
  non-trivial copoints:
  \begin{itemize}[label=-]
  \item $4 < 3 < 18 < 8 < 16$,
  \item $13 < 14 < 11$,
  \item $19 < 5 < 15 < 2 < 12$,
  \end{itemize}
  None of these copoints is separable, hence we obtain the points
  $17,9,10,6,4,13,7,19$ as representatives of the quotient space, on
  which we call $\variable{sortByBipartition}$ with $p = 1$ on
  Line~\ref{step:call-bi-partition}. As seen at the end of
  \Cref{subs:co-from-proxor}, this outputs the compatible order
  $19 < 13 < 4 < 9 < 1 < 17 < 10 < 6 < 7$. Then on
  Line~\ref{step:return} concatenating all the copoints in the same
  order as their representatives, we get the compatible order:
  $$ 19 < 5 < 15 < 2 < 12 < 13 < 14 < 11 < 4 < 3 < 18 < 8 < 16 < 9 < 1 < 17 < 10 < 6 < 7.$$
  We check that this order is compatible by verifying the monotonicity
  of rows and columns of \Cref{TABLE_gros_example_compatible}.

  To illustrate separable copoints, consider the
  recursive call to the copoint $\{2,5,12,15,19\}$. With $p = 2$ as
  pivot, the ordered copoint partition is
  $[\{2\}, \{15\}, \{5,12,19\}]$. We determine recursively that
  $\{5,12,19\}$ has $19 < 5 < 12$ as a compatible order. As seen 
  after \Cref{def:separable}, $\{5,12,19\}$ is a separable copoint
  with halved copoints $\{5,19\}$ and $\{12\}$, thus we can take
  $[2,15,19,12]$ as extended quotient space. Then
  $\variable{sortByBipartition}$ returns the order $19 < 15 < 2 < 12$,
  from which we get the compatible order $19 < 5 < 15 < 2 < 12$.
\end{runningexample}

\begin{figure}[htb]
  \scriptsize
  \begin{equation*}
    \begin{array}{cp{0.1cm}ccccccccccccccccccc}
      & & 19 &  5 & 15 &  2 & 12 & 13 & 14 & 11 &  4 &  3 & 18 &  8 & 16 &  9 &  1 & 17 & 10 &  6 &  7 \vspace{0.1cm}\\
      19 & &  0 &  1 &  2 &  2 &  4 &  5 &  5 &  5 &  8 &  8 &  8 &  8 &  8 & 10 & 10 & 10 & 10 & 10 & 11 \\
      5 & &    &  0 &  2 &  2 &  3 &  5 &  5 &  5 &  8 &  8 &  8 &  8 &  8 & 10 & 10 & 10 & 10 & 10 & 11 \\
      15 & &    &    &  0 &  1 &  2 &  5 &  5 &  5 &  8 &  8 &  8 &  8 &  8 & 10 & 10 & 10 & 10 & 10 & 11 \\
      2 & &    &    &    &  0 &  2 &  5 &  5 &  5 &  8 &  8 &  8 &  8 &  8 & 10 & 10 & 10 & 10 & 10 & 11 \\
      12 & &    &    &    &    &  0 &  5 &  5 &  5 &  8 &  8 &  8 &  8 &  8 & 10 & 10 & 10 & 10 & 10 & 11 \\
      13 & &    &    &    &    &    &  0 &  1 &  1 &  6 &  6 &  6 &  6 &  6 &  9 &  9 &  9 &  9 &  9 &  9 \\
      14 & &    &    &    &    &    &    &  0 &  1 &  6 &  6 &  6 &  6 &  6 &  9 &  9 &  9 &  9 &  9 &  9 \\
      11 & &    &    &    &    &    &    &    &  0 &  6 &  6 &  6 &  6 &  6 &  9 &  9 &  9 &  9 &  9 &  9 \\
      4 & &    &    &    &    &    &    &    &    &  0 &  1 &  2 &  2 &  3 &  9 &  9 &  9 &  9 &  9 &  9 \\
      3 & &    &    &    &    &    &    &    &    &    &  0 &  2 &  2 &  2 &  9 &  9 &  9 &  9 &  9 &  9 \\
      18 & &    &    &    &    &    &    &    &    &    &    &  0 &  2 &  2 &  9 &  9 &  9 &  9 &  9 &  9 \\
      8 & &    &    &    &    &    &    &    &    &    &    &    &  0 &  2 &  9 &  9 &  9 &  9 &  9 &  9 \\
      16 & &    &    &    &    &    &    &    &    &    &    &    &    &  0 &  9 &  9 &  9 &  9 &  9 &  9 \\
      9 & &    &    &    &    &    &    &    &    &    &    &    &    &    &  0 &  4 &  6 &  8 &  9 &  9 \\
      1 & &    &    &    &    &    &    &    &    &    &    &    &    &    &    &  0 &  4 &  8 &  8 &  9 \\
      17 & &    &    &    &    &    &    &    &    &    &    &    &    &    &    &    &  0 &  7 &  8 &  9 \\
      10 & &    &    &    &    &    &    &    &    &    &    &    &    &    &    &    &    &  0 &  7 &  9 \\
      6 & &    &    &    &    &    &    &    &    &    &    &    &    &    &    &    &    &    &  0 &  9 \\
      7 & &    &    &    &    &    &    &    &    &    &    &    &    &    &    &    &    &    &    &  0 \\
    \end{array}
  \end{equation*}
  \caption{The distance matrix $D_C$ of the Robinson space of
    \Cref{TABLE_gros_example} with the entries ordered along the
    compatible order
    $19, 5, 15, 2, 12, 13, 14, 11, 4, 3, 18, 8, 16, 9, 1, 17, 10, 6, 7$.
  }
  \label{TABLE_gros_example_compatible}
\end{figure}

\section{Conclusion}\label{s:conclusion}

In this paper, we investigated the structure of mmodules and copoint
partitions in general dissimilarity spaces, and, more particularly, in
Robinson spaces. We proved that the mmodules of any dissimilarity
space can be represented using the mmodule-tree and that the maximal
mmodules not containing a given point form a partition (which we
called a copoint partition). The copoint partition leads to the
quotient dissimilarity space, which reflect the large scale structure
of the dissimilarity space. In Robinson spaces, we proved that the
mmodules and the copoint partitions satisfy stronger properties. We
classified the copoints into separable, non-separable, and tight and
proved that in any compatible order each separable copoint define two
intervals, each non-separable copoint is a single interval, and that
there exists a compatible order in which all tight copoints are
intervals. After partitioning each separable copoint into two parts,
we obtain the extended quotient space. We prove that any such extended
quotient admits a $p$-proximity order and we show how to compute it
efficiently.

Based on all these results and notions, we presented a
divide-and-conquer algorithm for recognizing Robinson matrices in
optimal $O(n^2)$ time. Our algorithm first computes a copoint
partition, then recursively computes a compatible order of each
copoint of the partition, classifies the copoints and partitions the
separable copoints, constructs a $p$-proximity order of the extended
quotient. Finally, from the $p$-proximity order and the compatible
orders of its copoints, it derives a compatible order of the whole
space. Our algorithm does not partition the tight copoints, although
there may exist compatible orders in which some tigh copoints are
partitioned. Thus, our algorithm does not return all compatible
orders. In the companion paper~\cite{CaChNaPr_PQ}, we also establish a
correspondence between the mmodule-tree of a Robinson dissimilarity
and its PQ-tree. PQ-trees are used to encode all compatible orders of
a Robinson space and using one such compatible order (say, computed by
our algorithm), it can be shown that one can construct the PQ-tree.

Mmodule-trees, the copoints partitions, and their quotient spaces may
be viewed as generic ingredients when investigating general
dissimilarity spaces and may be useful for the recognition of other
classes of dissimilarities, in particular of tree-Robinson and
circular-Robinson dissimilarities. For example, the approach via
mmodules was one of the starting points of our optimal algorithm for
strict circular seriation in~\cite{CaChNaPr_circ}. One can also easily
characterize ultrametrics by their mmodule-tree.

As we already mentioned in the introduction, the first recognition
algorithm of Robinson spaces running in optimal $O(n^2)$ time was
presented by Préa and Fortin~\cite{Prea}. It first constructs (by the
Booth and Lueker's algorithm~\cite{BoothLueker}) a PQ-tree
representing a super set of the compatible permutations (if the
dissimilarity is Robinson). In a second stage, the PQ-tree is refined,
in $O(n^2)$ time, in a such way that the set of represented
permutations coincides with the set of the compatible ones. The Booth
and Lueker's algorithm is renowned to be tricky to be efficiently
implemented, and the second step is even more evolved. Due to this,
even if optimal, the algorithm of \cite{Prea} is not simple.


Our optimal recognition algorithm is simple and was relatively easy to
implement in OCaml, as it should also be in any mainstream programming
language. Since it uses only basic data structures, it can be casted
as practical. The program solves seemingly hard instances on 1000
points in half a second, and instances on 10000 points in less than a
minute on a standard laptop (the algorithm of~\cite{Prea} was
implemented by Préa and does not show the same practical
performances). Among the different random generators we used to
evaluate our program, the hardest instances were obtained by shuffling
Robinson Toeplitz matrices with coefficients in $\{0,1,2\}$.

\subsection*{Acknowledgement}
We would like to acknowledge the referees for their careful reading of
the manuscript and useful suggestions and comments. This research was
supported in part by {\sc anr} project {\sc distancia} ({\small{ANR-17-CE40-0015}}) and has
received funding from Excellence Initiative of Aix-Marseille - {\sc a*midex
}(Archimedes Institute {\small{AMX-19-IET-009}}), a French ``Investissements
d’Avenir” Program.



\newpage
{
\bibliographystyle{siamplain}
\bibliography{citations1}
}

\end{document}